\documentclass[fleqn,usenatbib,useAMS,twocolumn]{mnras}
\usepackage{graphicx}	
\usepackage{amsmath}	
\usepackage{amssymb}	
\usepackage{multicol}        
\usepackage{bm}		
\usepackage{pdflscape}	
\usepackage[T1]{fontenc}
\usepackage{ae,aecompl}
\usepackage{newtxtext,newtxmath}
\usepackage{subfig}
\usepackage{caption}
\usepackage{multicol}
\usepackage{subcaption}
\usepackage{epstopdf}
\title{Search for Quasi-Periodic Oscillations in \textsl{TESS} light curves of bright Fermi Blazars}

\author[Tripathi et al.]{
Ashutosh Tripathi,$^{1,2}$\thanks{E-mail: ashutosht@tamu.edu}
Krista Lynne Smith,$^{1,2}$
Paul J. Wiita,$^{3}$
and Robert V. Wagoner$^{4}$
\\
$^{1}$George P. and Cynthia Woods Mitchell Institute for Fundamental Physics and Astronomy, Texas A\&M University, College Station, TX 77843-4242, USA \\
$^{2}$Department of Physics, Southern Methodist University, 3215 Daniel Avenue, 
Dallas, Texas 75205, USA\\
$^{3}$Department of Physics, The College of New Jersey, 2000 Pennington Rd., Ewing, New Jersey 08628-0718, USA\\
$^{4}$Department of Physics and KIPAC, Stanford University, Stanford, California 94305, USA
}

\pubyear{2023}
\begin{document}
\label{firstpage}
\pagerange{\pageref{firstpage}--\pageref{lastpage}}
\maketitle

\begin{abstract}
In a previous paper, we reported evidence for quasi-periodicities in the \textsl{TESS} light curves of  BL Lacerate and two other blazars found serendipitously in the SDSS AGN catalog. In this work, we find tentative evidence for quasi-periodic features in the \textsl{TESS} observations of five sources in the fourth catalog of the Fermi--LAT (4FGL) sources:  J090453.4$-$573503, J2345$-$1555,  B0422+004, J002159.2$-$514028, and B0537$-$441.  We analysed the \textsl{TESS} light curves of these blazars that we extracted using a customized approach. The quasi-periodic oscillations (QPOs) are searched for using two timing analysis techniques: generalized Lomb-Scargle periodogram and weighted wavelet Z-transform.  Their apparent periods lie in the range of 2.8--6.5 days and have at least 3$\sigma$ significance in both of these methods.  QPOs at such timescales can originate from the kink instability model which relates the quasi-periodic feature with the growth of kinks in the magnetized relativistic jets. We performed MCMC simulations to obtain the posterior distribution of parameters associated with this model and found the kink period consistent with previous studies. 
\end{abstract}


\section{Introduction}

Supermassive black holes (SMBH) are believed to reside at the centers of galaxies and have masses between $10^5$--$10^{10}$ M$_\odot$. The gravity of such a massive central compact object can lead to accretion from its surroundings so that the central region becomes highly luminous and sometimes even outshines the whole galaxy. Such a system is referred to as an active galactic nucleus (AGN) which is classified further based on different properties including optical spectroscopy, inclination towards the line of sight, and radio properties \citep[e.g.][]{1989AJ.....98.1195K, 1995PASP..107..803U}. Blazars are radio-loud AGNs having jets pointed toward the line of sight, thereby amplifying the observed brightness through relativistic effects. Blazars are further classified into BL Lacertae objects and flat spectrum radio quasars (FSRQs). They exhibit extreme variability throughout the whole electromagnetic spectrum, from radio to $\gamma$ rays \citep[see][and references therein]{2008Natur.452..966M}. 


Quasi-periodic oscillations (QPOs) have been detected rarely in AGN but quite frequently in stellar-mass black hole and neutron star binary systems \citep{2006ARA&A..44...49R}. The timescales of reported QPOs in AGNs come from observations taken throughout the entire electromagnetic spectrum and range from a few tens of minutes to weeks and even years \citep[see][and references therein]{2003ApJ...585..665H, 2008Natur.455..369G, 2008ApJ...679..182E, 2013MNRAS.436L.114K, 2015Natur.518...74G, 2018ApJ...860L..10S, 2019MNRAS.484.5785G, 2021MNRAS.501.5997T, 2022Natur.609..265J, 2023T}. 

These QPO features could originate in the vicinity of the black hole and thus might be used to study the effect of gravity of the central object on its surroundings. Several models have been proposed to explain the origin of such features, including diskoseismology, warped accretion discs, and disc-jet coupling \citep{1999ApJ...524L..63S, 1999A&A...349.1003T, 2001ApJ...559L..25W, 2005PASJ...57..699K, 2009MNRAS.397L.101I}. Recently, \citet{2022ApJ...933...55C} performed three-dimensional general relativistic magnetic hydrodynamical simulations of accretion flows and argued that quasi-periodic episodic jets could arise from flux ropes formed via magnetic reconnection in the flows. The frequencies of QPOs associated with accretion discs scale inversely with the mass of the central object. Interestingly, this relation is valid for both stellar mass and many AGN black holes and suggests that the origin of these periodicities could be the same in all classes of black hole-powered sources \citep{2004ApJ...609L..63A}.

The Transiting Exoplanet Survey Satellite (\textsl{TESS}) \citep{2015ESS.....350301R} is a space-based optical instrument. Unlike temporal studies made with ground-based telescopes,  \textsl{TESS} is not affected by seasonal gaps, irregular sampling, and  degraded photometric precision arising from the atmosphere. The variability properties of a few blazars have already  been studied using \textsl{TESS} observations \citep{2020ApJ...900..137W, 2021MNRAS.501.1100R}. Recently,  we \citep{2023T} claimed the presence of quasi-periodicity in \textsl{TESS} light curves of BL Lacertae itself and two other blazars discovered serendipitously  in the SDSS-IV SPIDERS catalog, which essentially consists of X-ray selected AGNs from  the ROSAT All-Sky Survey (RASS). 

In section \ref{sec:1} we explain the sample selection and custom \textsl{TESS} data reduction. Data analysis methods are explained in section \ref{sec:2a}. The timing analysis results are discussed in section \ref{sec:2}.  We discuss some theoretical models that could explain the claimed QPO periods in section \ref{sec:3}. Markov Chain Monte Carlo (MCMC) simulations of the kink instability model are discussed in Appendix \ref{app:2}.

\section{Sample Selection and Data Reduction}\label{sec:1}

\subsection{\textsl{TESS} Monitoring and Blazar Sample}
The Transiting Exoplanet Survey Satellite \citep[\textsl{TESS}; ][]{2015ESS.....350301R} conducts long-term monitoring with high photometric precision of stars across nearly the entire sky with a rapid cadence: every 30 minutes in the early cycles (2018 -- 2019) and every 10 minutes or 2 minutes in later and current cycles (2020 -- present). Except for a roughly 1-day long gap every orbit (approximately every two weeks), the coverage is continuous. The monitoring baseline for a \textsl{TESS} source depends upon its ecliptic latitude. The majority of sources at low-to-middle ecliptic latitudes are monitored for 27 days before the spacecraft shifts to the next wedge of the sky. Each 27-day monitoring segment is referred to as a ``Sector." At higher ecliptic latitudes, these sectors overlap, resulting in a longer monitoring baseline, with a maximum of approximately 1 year of continuous monitoring at the ecliptic poles. After this year of exposure, the spacecraft turns over and surveys the other ecliptic hemisphere for 1 year with the same pattern. 

To determine our sample for this analysis, we selected all objects from the fourth Fermi Large Area Telescope catalog of $\gamma$-ray sources \citep[4FGL; ][]{2020ApJS..247...33A} that are classified as BL Lac, FSRQ, or ``unknown blazar type." We then cross-matched this list with the \textsl{TESS} Input Catalog \citep[TIC; ][]{2019AJ....158..138S}. The photometric precision of \textsl{TESS} is exquisite for bright sources, officially diminishing to 1\% for sources with $V\sim16$ in pre-launch specifications (although actual performance is often better)\footnote{See the technical specs here: \url{https://heasarc.gsfc.nasa.gov/docs/tess/observing-technical.html}}. We, therefore, require that the \textsl{TESS} Magnitude (i.e., the magnitude in the \textsl{TESS} bandpass as reported by the TIC) be $<17.5$. This cross-match and magnitude cut resulted in a total sample of 156 blazars.

\subsection{Light Curve Extraction with \texttt{Quaver}}
The light curve extraction and systematics correction were done using the \texttt{Quaver} software, specially designed with \textsl{TESS} AGN science in mind. In particular, the code allows the user to custom-select their extraction aperture from a cutout of the full FFI without downloading the entire image, which assists in avoiding nearby contaminating sources and can accommodate extended host galaxies. This is achieved using the TESSCut package \citep{2019ASPC..523..397B}. The program then makes use of several Lightkurve \citep{lightkurve} tasks. Standard principal component analysis of faint background pixels and background-corrected bright source pixels (excluding the target) is used to build design matrices housing the components of the additive background light and the more subtle multiplicative systematics affecting sources. These matrices are then used by Lightkurve's \texttt{RegressionCorrector} to correct the target's light curve. The approach is flexible, allowing the user to determine the number of principal components used, and whether to handle the background using regression or simple subtraction. The \texttt{Quaver} code \citep{2023ApJ...958..188S} is publicly available, with a detailed User Guide describing the different extraction modes\footnote{\url{https://github.com/kristalynnesmith/quaver}}. 

Of the 156 blazars in our sample, 15 were excluded due to excessive crowding in the source region (usually at low Galactic latitudes), which precluded isolating the counterpart on \textsl{TESS}'s large 21\arcsec pixels. Of the remaining sample, 44 sources are found to be variable from visual inspection. Of this subset, 12 sources are found to have indications of periodicity in their light curves supported by one or more different timing analysis techniques. Finally, only 5 sources ($\approx 3$\%) appear to have  QPOs that survived our significance tests.  Four of these light curves show at least 3~$\sigma$ significance in both of the analysis approaches we used and the other one has such high significance in one of those methods.

    \begin{figure*}
    \centering
    \bf{ (a) ATPMN~J090453.4573503~Sec.~37 }\hspace{6.0cm}\bf{(b) QSO J2345-1555 Sec. 29}
        \centering
        \begin{multicols}{2}
            \vspace*{-0.5cm}{\includegraphics[width=\linewidth]{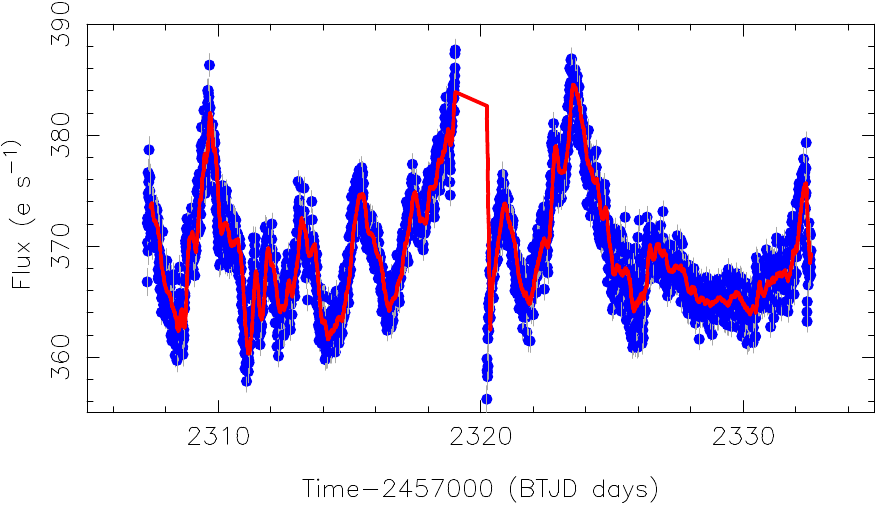}}\par 
            {\includegraphics[width=\linewidth]{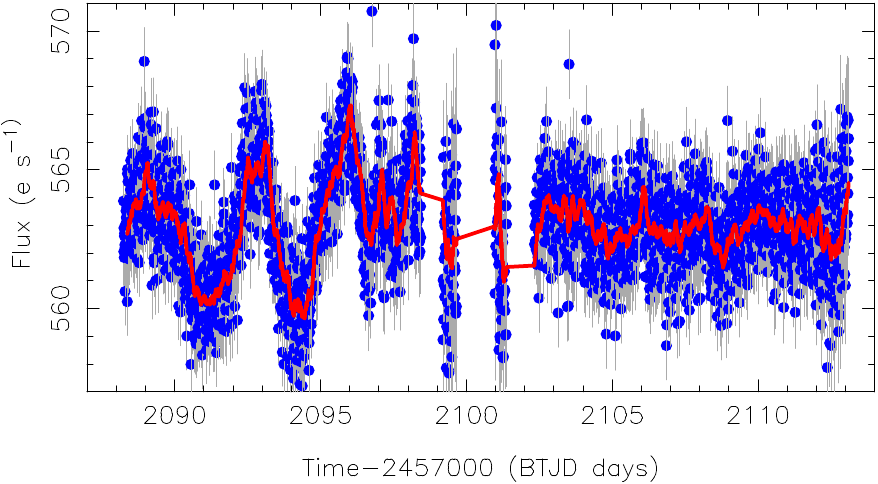}}\par 
        \end{multicols}
        \vspace{-0.9cm}
        \begin{multicols}{2}
            {\includegraphics[width=\linewidth]{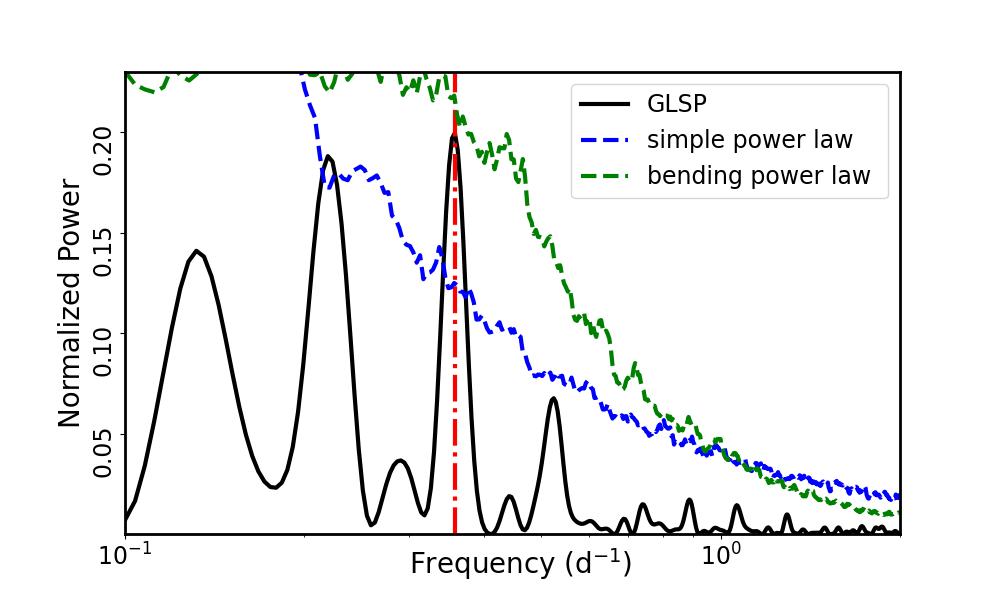}}\par
            {\includegraphics[width=\linewidth]{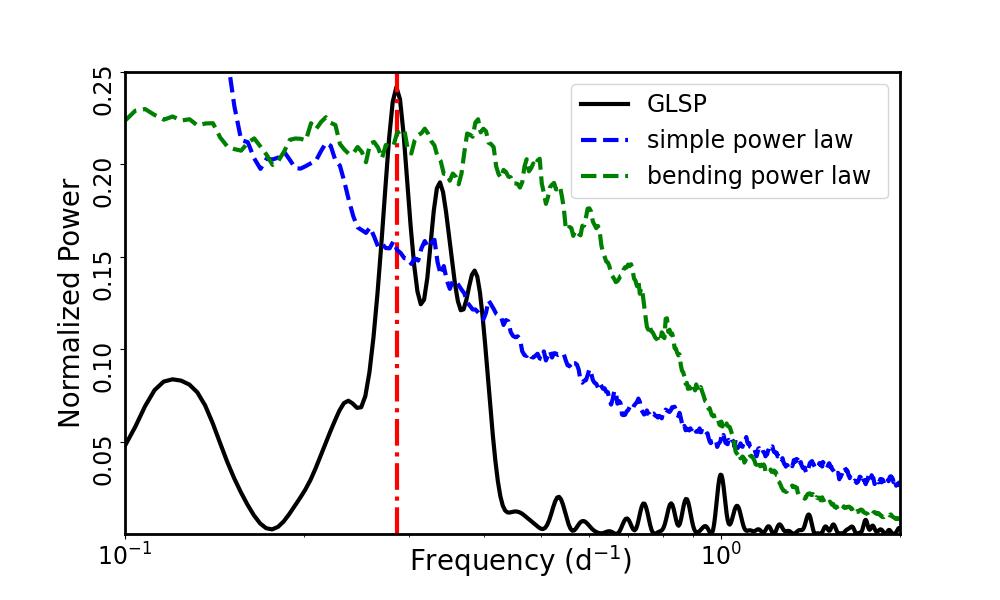}}\par
        \end{multicols}
        \vspace{-1.8cm}
        \begin{multicols}{2}
            {\includegraphics[angle=90, width=\linewidth]{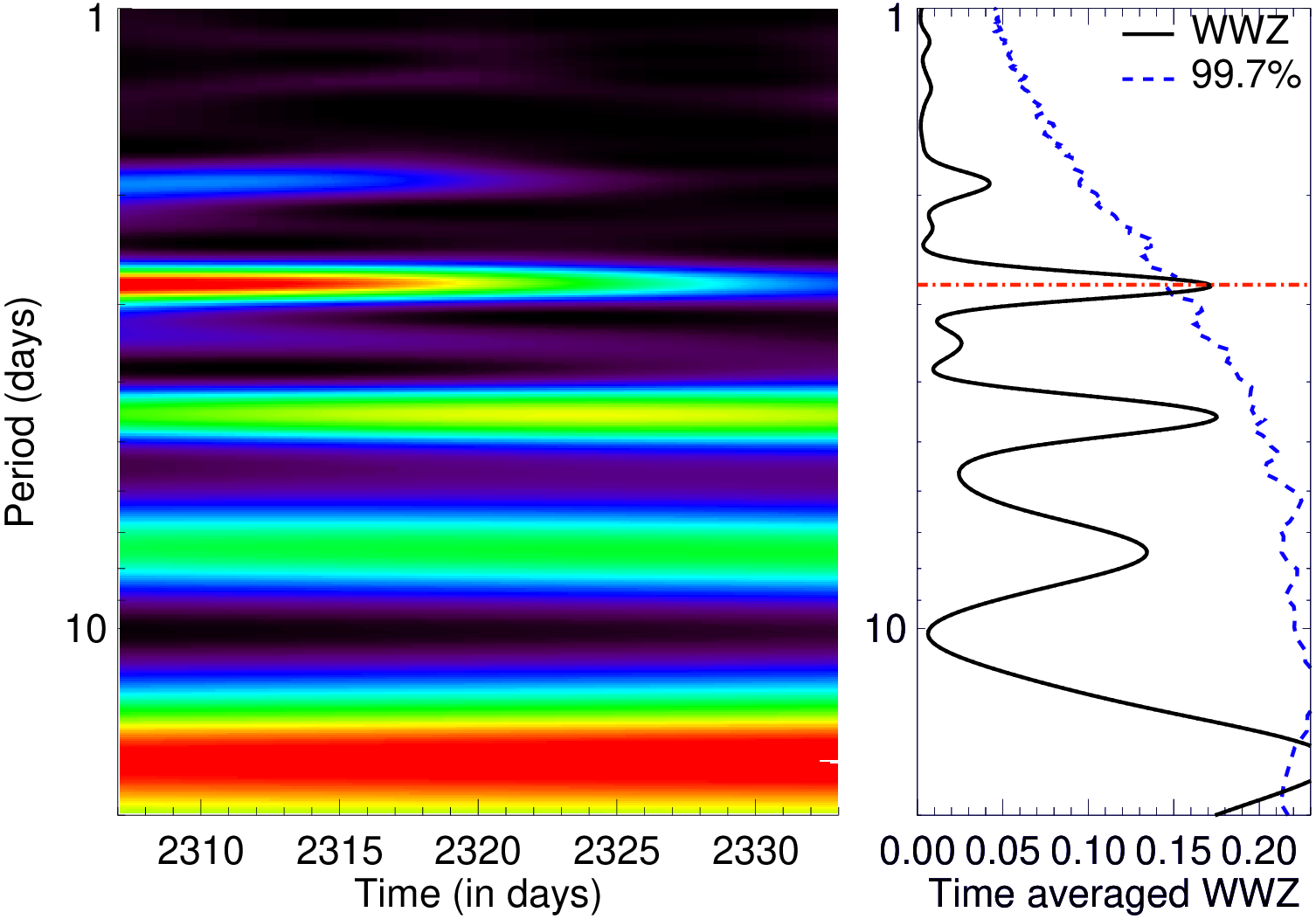}}\par
            {\includegraphics[angle=90, width=\linewidth]{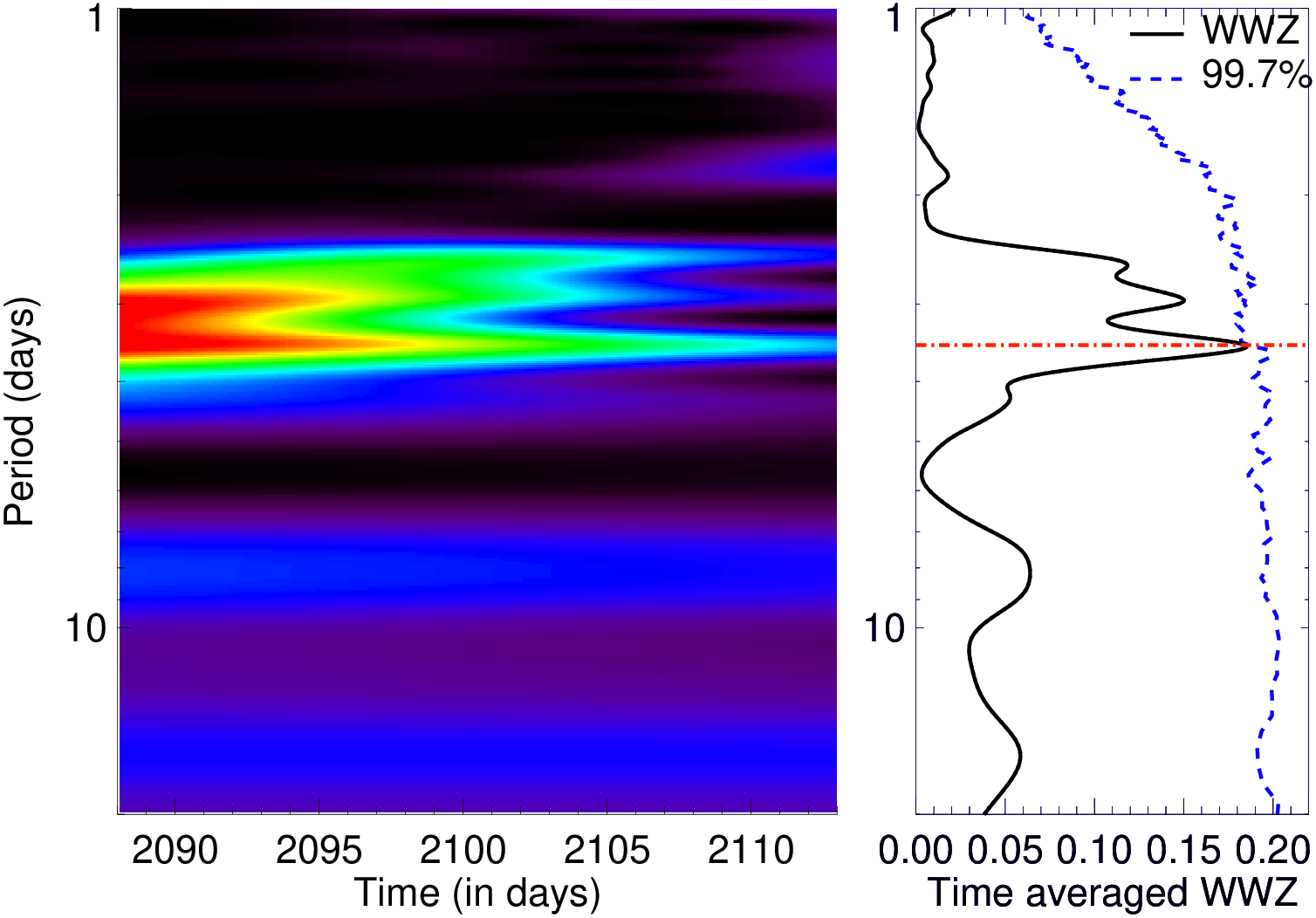}}\par
        \end{multicols}
    \caption{Results for (a) ATPMN J090453.4$-$573503 during \textsl{TESS} sector 37 and (b) QSO J2345$-$1555 during sector 29.  The upper panels show the light curves analysed in this work for these two sources, where blue dots show the actual data points, gray denotes the associated error, and the red curve corresponds to the running average. The middle panel for each source shows a generalized Lomb-Scargle periodogram (GLSP). The left plot of the bottom panel for each source shows the weighted wavelet Z-transform analysis and the right plot shows the time-averaged WWZ power. The red dotted-dashed curves in the GLSP and time-averaged WWZ plots denote the claimed QPO signal. Blue dashed curves represent the 99.73\% (3~$\sigma$) global significances, while the green dashed curve in the GLSP results corresponds to the power spectrum obtained from a bending power-law fit.}
    \end{figure*}



    \begin{figure*}
    \centering
    \bf{ (a) QSO B0422+004 Sec. 5 }\hspace{6.0cm}\bf{(b) QSO B0422+004 Sec. 32}
        \centering
        \begin{multicols}{2}
            {\includegraphics[width=\linewidth]{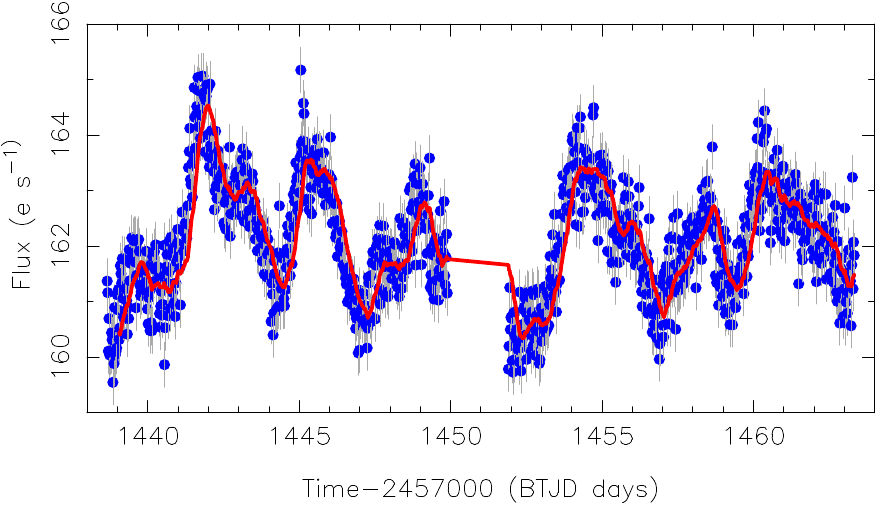}}\par 
            {\includegraphics[width=\linewidth]{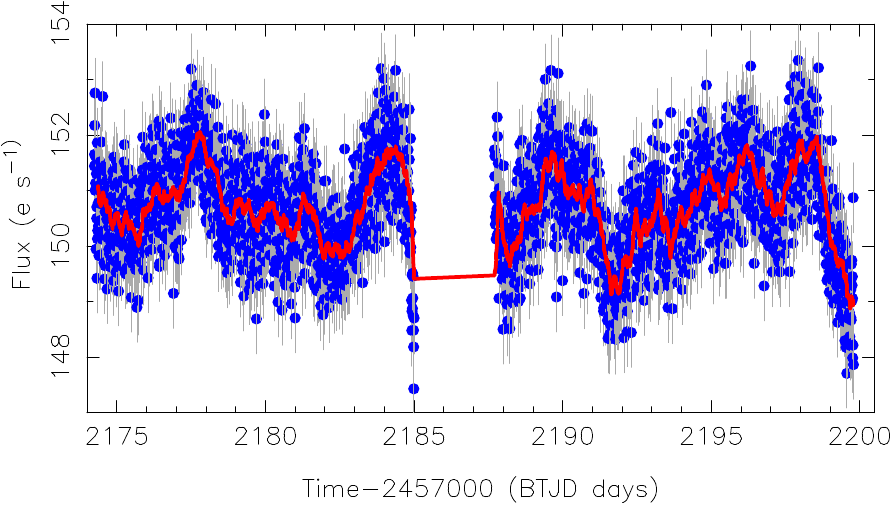}}\par 
        \end{multicols}
        \vspace{-0.9cm}
        \begin{multicols}{2}
            {\includegraphics[width=\linewidth]{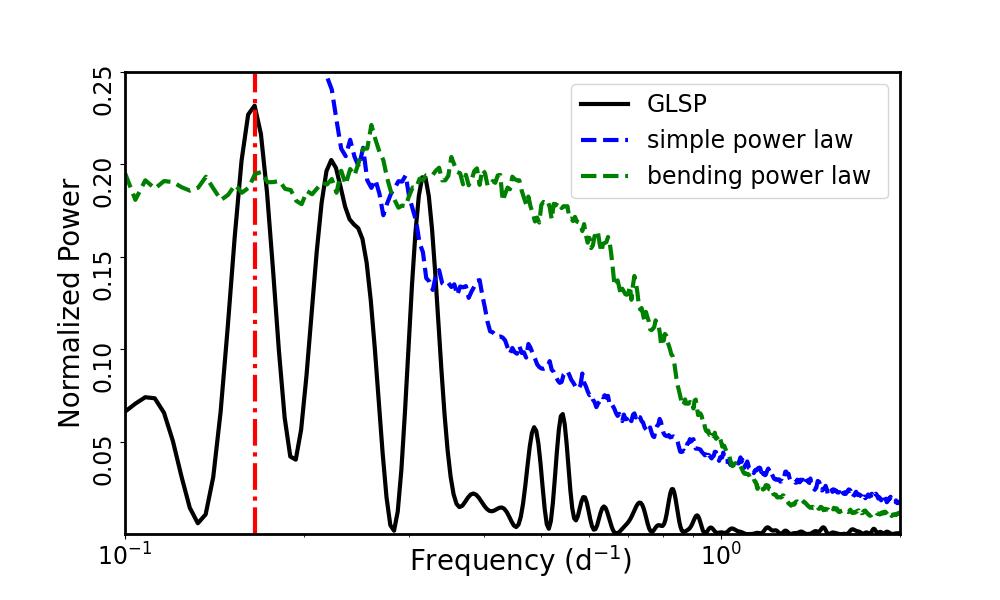}}\par
            {\includegraphics[width=\linewidth]{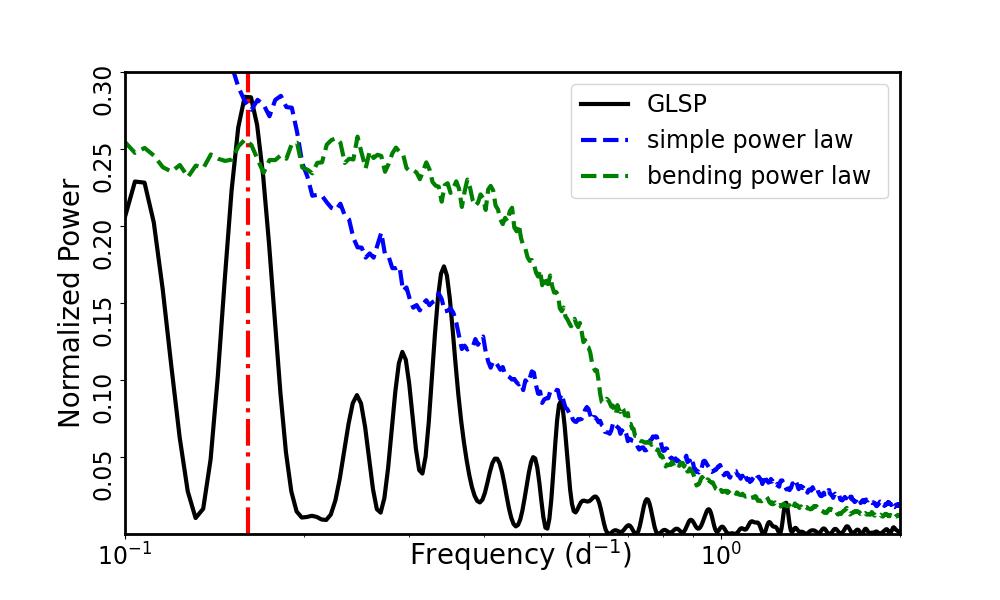}}\par
        \end{multicols}
        \vspace{-1.8cm}
        \begin{multicols}{2}
            {\includegraphics[angle=90, width=\linewidth]{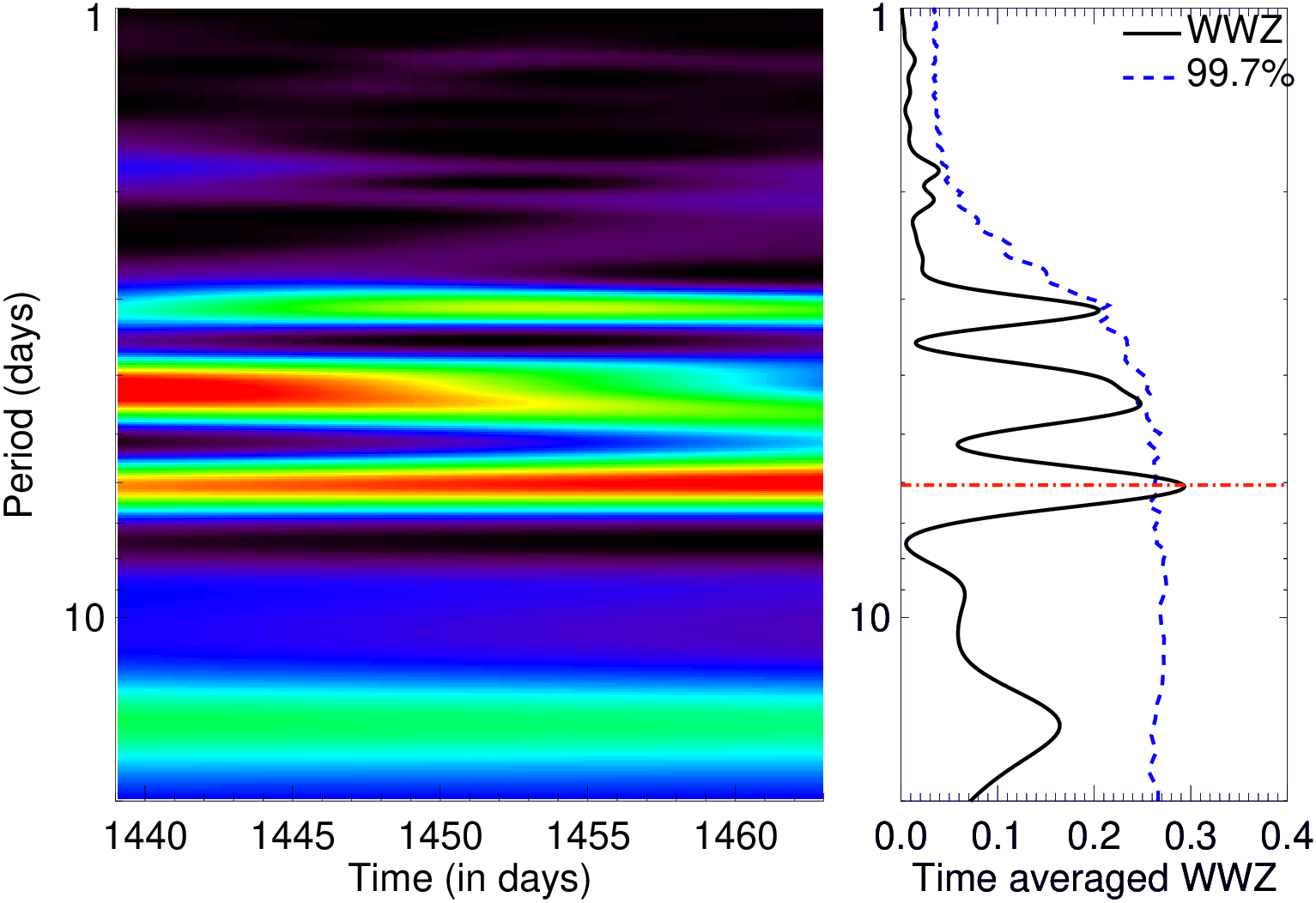}}\par
            {\includegraphics[angle=90, width=\linewidth]{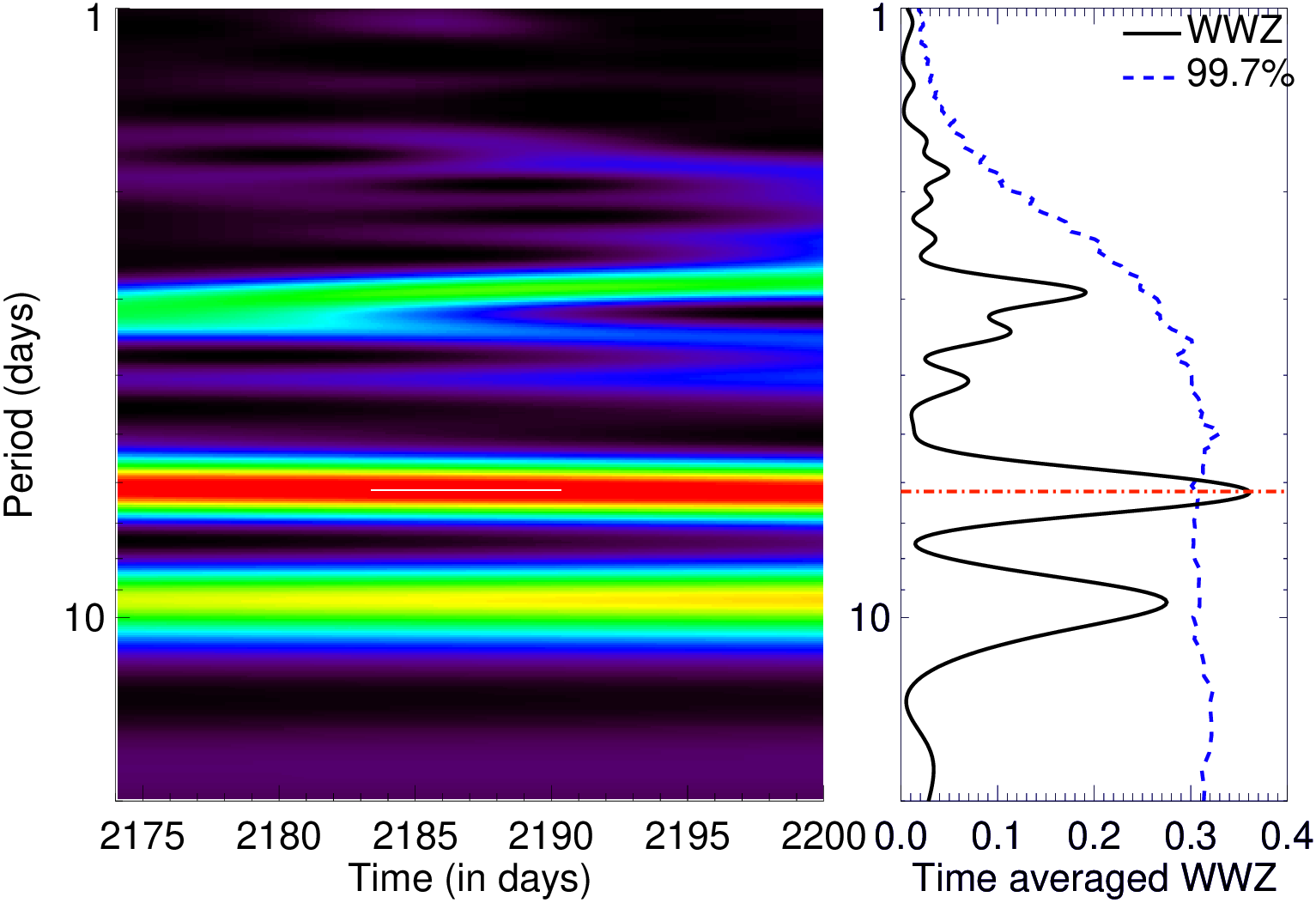}}\par
        \end{multicols}
    \caption{Same as in Fig.~1 but for Sec.~5 (left) and Sec.~32 of QSO B0422+004.}
    \end{figure*}

    \begin{figure*}
    \centering
    \bf{ (a) 1RXS J002159.2-514028 Sec.~2 }\hspace{6.0cm}\bf{(b) QSO B0537-441 Sec.~ 32 \& 33}
        \centering
        \begin{multicols}{2}
            \vspace*{-0.57cm}{\includegraphics[width=\linewidth]{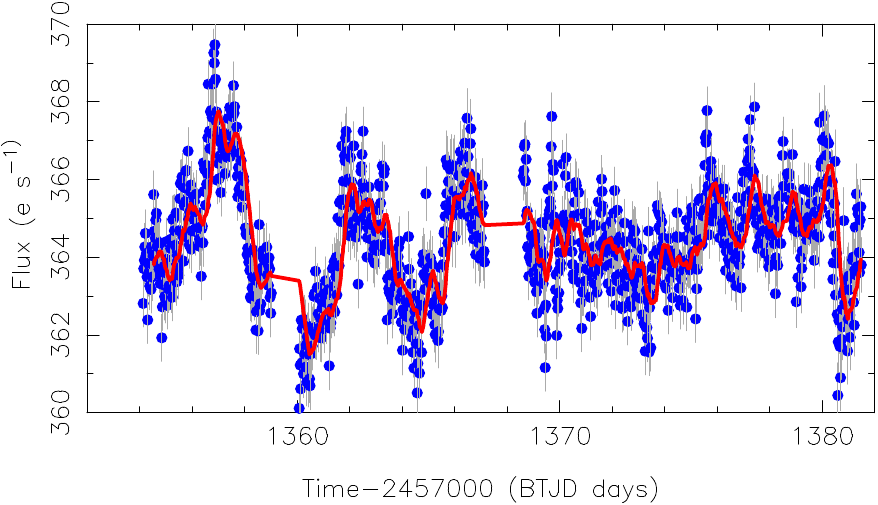}}\par 
            {\includegraphics[width=\linewidth]{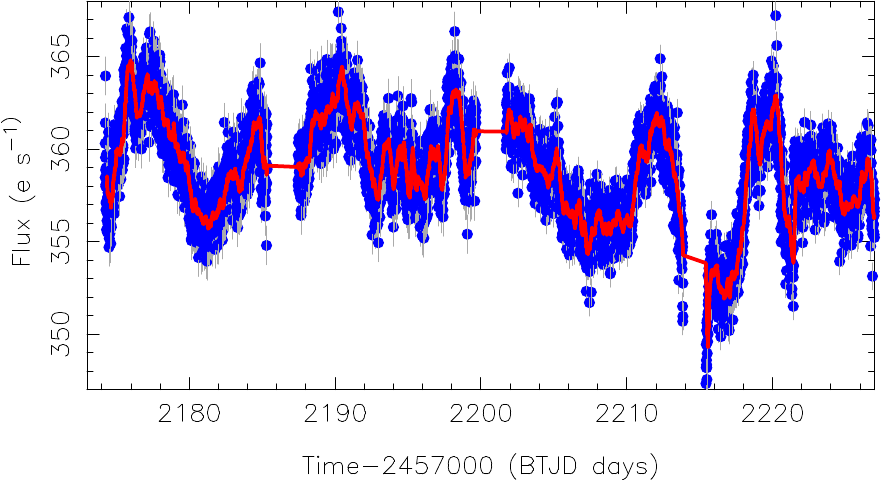}}\par 
        \end{multicols}
        \vspace{-0.9cm}
        \begin{multicols}{2}
            {\includegraphics[width=\linewidth]{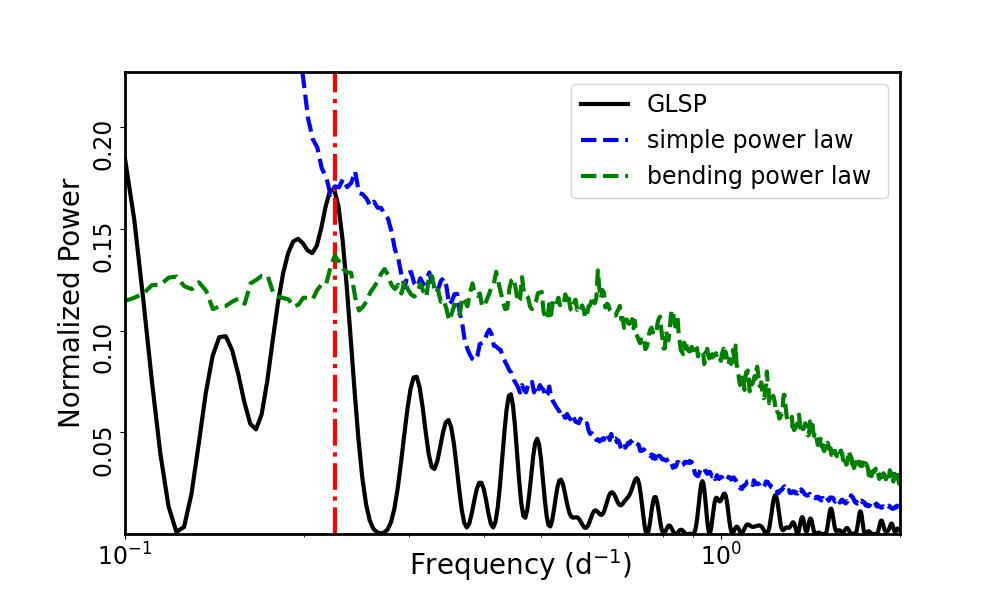}}\par
            {\includegraphics[width=\linewidth]{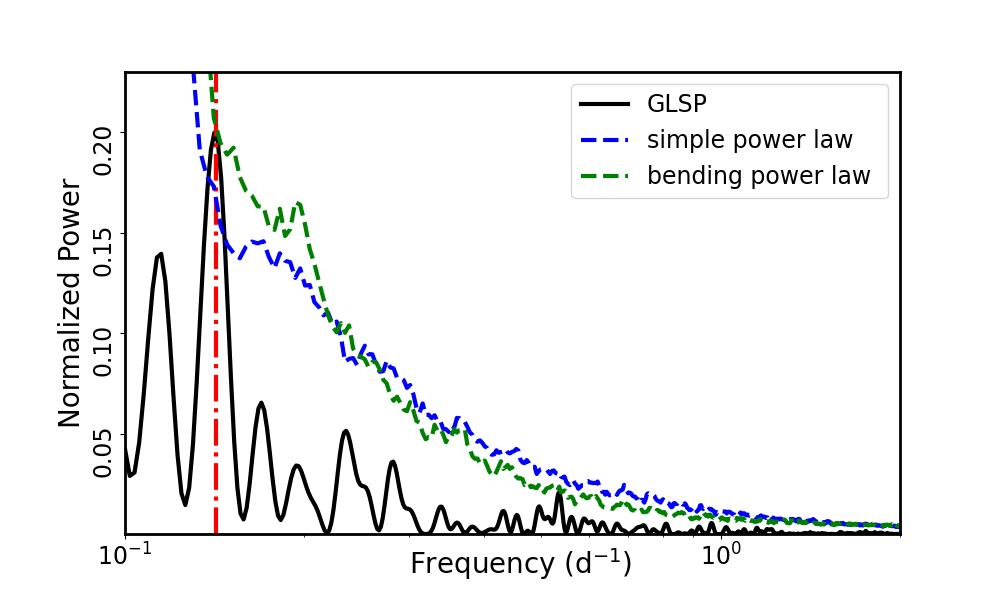}}\par
        \end{multicols}
        \vspace{-1.8cm}
        \begin{multicols}{2}
            {\includegraphics[angle=90, width=\linewidth]{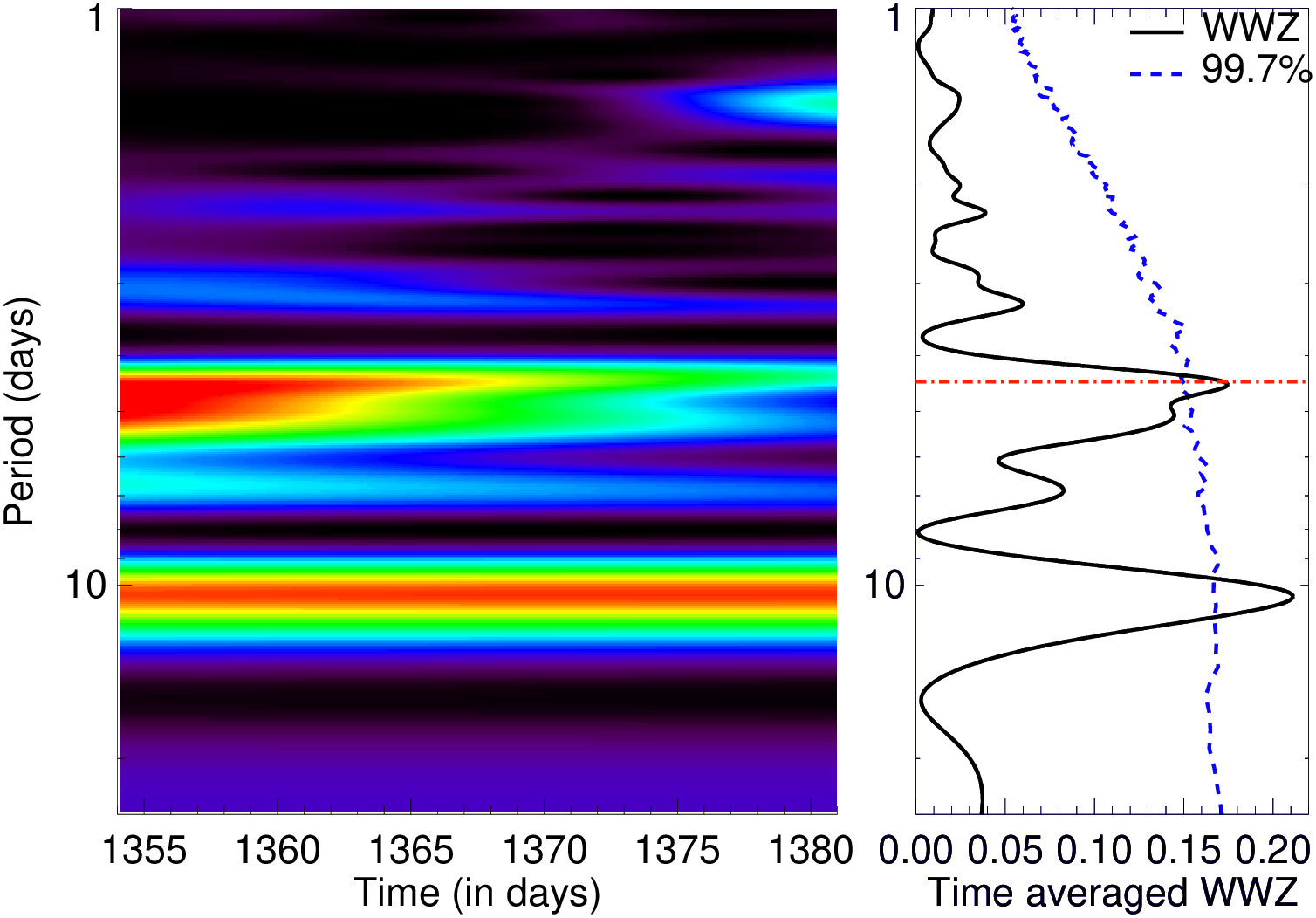}}\par
            {\includegraphics[angle=90, width=\linewidth]{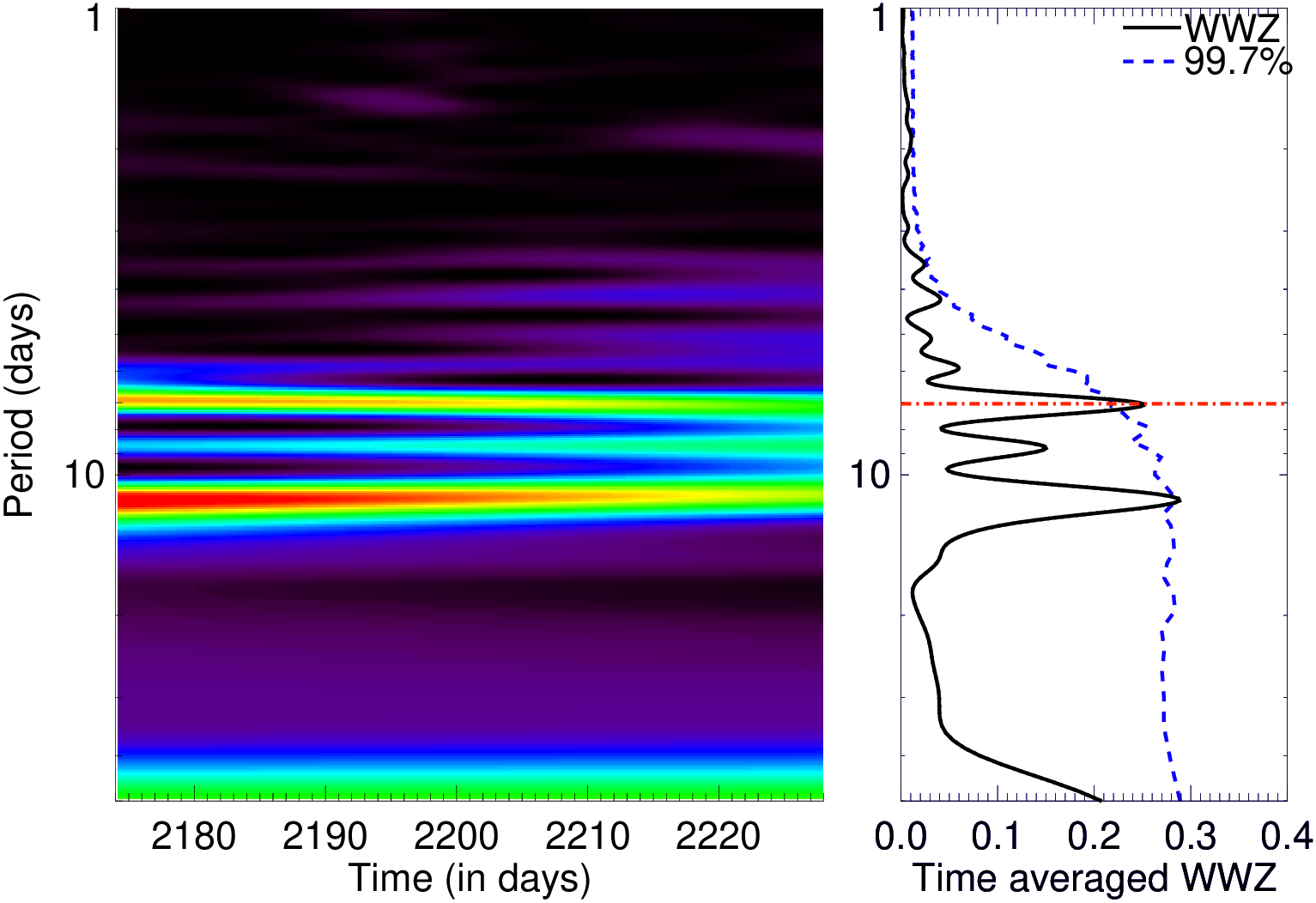}}\par
        \end{multicols}
    \caption{Same as in Fig.~1 but for  (a) 1RXS J002159.2-514028 Sec.~2 and (b) QSO B0537-441 Sec.~ 32 \& 33.}
    \end{figure*}

\section{Data analysis methods}\label{sec:2a}

\paragraph*{Generalized Lomb Scargle Periodogram (GLSP)} The periodogram of a given time series
can be estimated by a discreet Fourier transform and used to search for periodicities. The Lomb-Scargle periodogram (LSP) is a commonly used Fourier transform method employed in analyzing irregular and unevenly sampled observations \citep{1976Ap&SS..39..447L, 1982ApJ...263..835S} by fitting a time series with sinusoidal components. The generalized LSP (GLSP) is a modified version of the LSP where not only sinusoidal and constant components are used to fit a given time series but errors in the fluxes are also included in the calculations \citep{2009A&A...496..577Z}. In this work, we use the GLSP code provided by {\tt pyastronomy}\footnote{\url{https://github.com/sczesla/PyAstronomy}}. The employed code uses Leahy normalization which includes the number of trials in calculations so that the significance is intrinsically globally estimated. Here, we set the oversampling factor to be 2.0 which means that the number of frequencies is equal to the number of data points in the observations. We also tried oversampling factors of 1.0 and 10.0 to assess their effects on the significance level. The significances calculated using these different oversampling factors are consistent within 1--2\% for all light curves analysed in this work. In this work, the analysed light curves are too short to make a clear determination of the underlying baseline model for the power spectrum. So we have employed  models which assume that the light curves of AGNs typically follow red noise due to stochastic processes in accretion discs or jets. 

We fit the  power spectrum of a given data both with a simple power law and with a bending power law  
using the {\tt scipy} python routine.  We then estimate the uncertainties on the parameters through MCMC simulations using 500 walkers and 200,000 iterations. The simple power law used is 
\begin{equation}
   P(\nu) = N \left(\frac{\nu}{\nu_0}\right)^{\beta},
\end{equation}
where $N$ is the normalization, $\nu$ is frequency, $\nu_0$ is a constant, and $\beta$ is the power-law index. The bending power law is expressed as 

\begin{equation}
   P = N \left(1+ \left(\frac{\nu}{\nu_b} \right)^{-\alpha}\right)^{-1.0} + C,
\end{equation}
where $\nu_b$ is the bending frequency,  $\alpha$ is the power-law index, and $C$ is the constant representing the instrumental white noise that dominates at the highest frequencies.


We then simulate 10,000 light curves, with the best-fit parameters and statistical properties of the observation obtained from the  timing analysis package {\tt stingray}\footnote{\url{https://github.com/StingraySoftware}} \citep{2019ApJ...881...39H} which uses an algorithm based on \citet{1995A&A...300..707T}. Each sector of \textsl{TESS} light curves have a gap in their middle due to the orbital motion of the instrument and have a duration in the range of $\sim$1--4 days. So, we simulate each light curve with a such a gap whose length is similar to that found in the real \textsl{TESS} observations. For QSO B0537$-$441, two consecutive sectors are analysed. In this case, we also include the sector gap in our calculation in addition to the gaps present in the middle of each sector's observations. We calculate GLSPs for all these simulated light curves and obtain the distribution of power spectra at individual frequencies. The spectral distribution at each frequency can be used to estimate the significance of a peak for a given confidence interval. For instance, the  99.73 \% confidence interval is estimated by the 99.73 percentile of the power distribution of these 10000 simulated light curves. 

\paragraph*{Weighted Wavelet Z-transform (WWZ)} A periodic signal may not persist throughout an observation. Wavelet analyses  allow for and assess the possible non-stationarity present in a given time series by decomposing the signal in the time and frequency domains simultaneously. WWZ is a commonly used wavelet technique used to analyse unevenly sampled and sparse time series \citep{2005NPGeo..12..345W}. We employed a WWZ code\footnote{\url{https://www.aavso.org/software-directory}} used commonly for such analysis  \citep[see][and references therein]{2017ApJ...849....9Z, 2021MNRAS.501.5997T, 2023T}. When we average the WWZ power over the whole period of duration, we get the time-averaged WWZ, which is a periodogram following a $\chi^2$ distribution with two degrees of 
freedom \citep{1996AJ....112.1709F}. As this time-averaged WWZ is a Fourier transform, we follow an approach similar to that we used for the GLSP to calculate any QPO's significance. See \citet{{2023T}} for more details.

\section{Results}\label{sec:2}

The upper panels of Figs.\ 1--3 show  the light curves produced using \texttt{Quaver} of the Fermi--LAT blazars analysed in this work. The red curve in these plots shows the running average of the observation obtained using {\tt window=20}, which essentially means the average of the present and the next 19 measurements is plotted, corresponding to smoothing over $\sim$10 hours duration. Visual inspection led us to note the possibility of  quasi-periodic behavior in each of these light curves and this was confirmed by the Generalized Lomb Scargle Periodogram (GLSP) and Weighted Wavelet Z-transform (WWZ) analysis methods.

Fig.~1(a) shows the timing analysis results obtained for the 25-day sector 37 (cycle 3) observation of ATPMN J090453.4$-$573503. A strong signal at a  period of around 2.8 days is at least 3$\sigma$ significant in the GLSP and WWZ results regardless of which of the red noise models is employed. The signal is found to be persistent throughout most of the observation as shown in the colour density (violet indicating the least power and red  the most) plot of WWZ power (left plot of bottom panel) which also shows the presence of non-stationarity in the observation. The time-averaged WWZ (right plot in bottom panel) supports the presence of the peak at the claimed frequency with 3$\sigma$ significance. There is also a peak around 4.5 days which is 3$\sigma$ only for the simple power law. This signal is also found in the WWZ analysis but has less power as compared to the peak around 2.8 days. 

Fig.~1(b) shows the results obtained for the 25-day sector 29 (cycle 3) observation of QSO J2345$-$1555. In the GLSP result, a strong peak having significance at least 3$\sigma$ is detected at 3.5 days assuming a either simple or a bending power law as the underlying model. The wavelet analysis shows this peak is entangled with other lower significance peaks and together, they are much stronger in the first half of the observation. In the time-averaged WWZ plot, this peak at 3.5 days has a significance of 3$\sigma$, so we accept it as meeting our full set of criteria or a tentative claim of a QPO.

The  light curves and the timing analysis results of QSO B0422+004 are plotted in Fig.~2. This source was observed during two different \textsl{TESS} cycles. The results for the 25-day sector 5 (cycle 1) observation are shown in Fig.~2(a). Using a bending power law, the peaks around 3.1, 4.5, and 6.0 days are found to be at least 3$\sigma$ significant, whereas with the simple power law, only the peak around 3.2 days is found to be significant. So this peak is at least 3$\sigma$ significant with both underlying models. The WWZ power of the 3.1-day peak is somewhat less than that of the peaks around 4.1 and 6.0 days. It persists throughout the observation with similar WWZ power; however, this is not the case for the 4-day signal which is strong only in the first half of the observation. In the time-averaged WWZ result, the peak around 6.0 days has more than 3$\sigma$. However since the length of the 25-day segment contains only  4 cycles of a putative 6-day QPO, it is difficult to make a strong claim for it  based solely on this sector. The 3.1-day peak is found to be at least 3$\sigma$ in both GLSP and WWZ methods and also corresponds to 8 cycles of a putative QPO. 
 The analysis  of the other 25-day sector 32 (cycle 3) data for this source (Fig.~2(b)), shows a strong (3$\sigma$) signal at 6.0 days in the GLSP with both red noise models. This is also present in time-averaged WWZ analysis with 3$\sigma$ significance, respectively. With a simple power law, a peak around 3 days is also found to be more than 3$\sigma$ significant.  The WWZ colour diagram indicates the persistence of this 6.3-day signal throughout the observation.  The fact that both the widely separated cycle 1 and cycle 3 observations indicate a QPO of around 6 days strengthens our confidence in its presence. The $\sim$6-day signal could be harmonically related to the signal detected at $\sim$3-day but because of the lower significance of the latter using both GLSP and WWZ methods, we take the former to be the most likely QPO period.

\begin{table*}
 \centering
 \caption{Best fit parameters and associated errors for the bending power-law sampled from MCMC realizations. }
\begin{tabular}{|c|c|c|c|c|c|}
\hline
Source \& Sector & Segment & log $N$ & $\alpha$ & $\nu_b$ & $C$\\\hline
ATPMN J090453.4$-$573503 Sec.\ 37 &1& $-0.91^{+0.22}_{-0.22}$& $-6.18^{+1.51}_{-1.52}$ & $0.45^{+0.11}_{-0.11}$ & $0.00112^{+0.0003}_{-0.0002}$ \\ 
 & 2 & $-1.88^{+0.44}_{-0.48}$& $-4.14^{+1.04}_{-1.03}$ & $0.96^{+0.24}_{-0.25}$ & $0.005^{+0.0001}_{-0.0001}$ \\ \hline

QSO J2345-1555 Sec.\ 29  & 1 & $-0.70^{+0.17}_{-0.18}$& $-4.51^{+1.11}_{-1.10}$ & $0.69^{+0.18}_{-0.17}$ & $0.002^{+0.0005}_{-0.0005}$ \\
 & 2 & $-1.26^{+0.32}_{-0.33}$& $-4.24^{+1.09}_{-1.04}$ & $0.84^{+0.20}_{-0.21}$ & $0.009^{+0.0023}_{-0.0023}$ \\ \hline
 
QSO B0422$+$004 Sec.\ 5 & 1 & $-1.32^{+0.34}_{-0.34}$& $-8.42^{+2.08}_{-2.14}$ & $0.71^{+0.17}_{-0.19}$ & $0.0009^{+0.0002}_{-0.0002}$ \\
 & 2 & $-1.95^{+0.49}_{-0.50}$& $-7.21^{+1.74}_{-1.76}$ & $0.93^{+0.23}_{-0.24}$ & $0.001^{+0.0003}_{-0.0002}$ \\ \hline
  QSO B0422$+$004 Sec.\ 32 & 1 & $-1.15^{+0.28}_{-0.29}$& $-3.47^{+0.91}_{-0.88}$ & $0.57^{+0.14}_{-0.15}$ & $0.002^{+0.0005}_{-0.0005}$ \\
 & 2 & $-1.03^{+0.27}_{-0.24}$& $-7.79^{+1.91}_{-1.94}$ & $0.53^{+0.13}_{-0.13}$ & $0.0032^{+0.0008}_{-0.0008}$ \\ \hline

1RXS J002159.2$-$514028 Sec.\ 2 & 1 & $-1.82^{+0.45}_{-0.44}$& $-6.13^{+1.52}_{-1.57}$ & $1.20^{+0.31}_{-0.31}$ & $0.0009^{+0.0002}_{-0.0002}$ \\
 & 2 & $-1.69^{+0.43}_{-0.41}$& $-1.75^{+0.46}_{-0.46}$ & $1.22^{+0.29}_{-0.30}$ & $0.0002^{+0.0001}_{-0.0001}$ \\ \hline

QSO B0537$-$441 Sec. 32 \& 33 & 1 & $-0.14^{+0.03}_{-0.03}$& $-8.92^{+2.34}_{-2.27}$ & $0.12^{+0.03}_{-0.03}$ & $0.006^{+0.0015}_{-0.0016}$ \\
 & 2 & $-0.13^{+0.03}_{-0.03}$& $-1.41^{+0.35}_{-0.35}$ & $0.08^{+0.02}_{-0.02}$ & $0.0002^{+0.0001}_{-0.0001}$ \\ 
  & 3 & $-0.04^{+0.01}_{-0.01}$& $-2.78^{+0.69}_{-0.73}$ & $0.09^{+0.02}_{-0.02}$ & $0.0012^{+0.0002}_{-0.0003}$ \\
   & 4 & $-2.24^{+0.55}_{-0.55}$& $-4.64^{+1.12}_{-1.11}$ & $1.29^{+0.32}_{-0.33}$ & $0.0009^{+0.0001}_{-0.0001}$ \\\hline

\end{tabular}
\end{table*}

 \begin{table*}
 \centering
 \caption{Best fit parameters and associated errors for the simple power-law sampled from MCMC realizations. }
\begin{tabular}{|c|c|c|c|c|}
\hline
Source \& Sector & Segment & log $N$ & $\beta$ & $\nu_0$ \\\hline

ATPMN J090453.4$-$573503 Sec.\ 37 &1& $-1.52^{+0.37}_{-0.38}$& $-1.04^{+0.27}_{-0.26}$ & $0.72^{+0.18}_{-0.18}$  \\ 
 & 2 & $-2.04^{+0.52}_{-0.47}$& $-1.35^{+0.35}_{-0.33}$ & $0.63^{+0.15}_{-0.16}$ \\ \hline

QSO J2345-1555 Sec.\ 29 & 1 & $-2.04^{+0.52}_{-0.51}$& $-1.08^{+0.25}_{-0.27}$ & $0.62^{+0.15}_{-0.16}$  \\
 & 2 & $-1.52^{+0.39}_{-0.37}$& $-0.56^{+0.14}_{-0.14}$ & $0.73^{+0.18}_{-0.18}$ \\ \hline

 QSO B0422$+$004 Sec.\ 5 & 1 & $-1.43^{+0.35}_{-0.34}$ & $-1.20^{+0.31}_{-0.30}$ & $0.63^{+0.15}_{-0.15}$  \\
 & 2 & $-1.99^{+0.49}_{-0.48}$& $-1.32^{+0.33}_{-0.32}$ & $0.65^{+0.16}_{-0.16}$ \\ \hline

QSO B0422$+$004 Sec.\ 32 & 1 & $-1.44^{+0.36}_{-0.36}$ & $-1.21^{+0.29}_{-0.31}$ & $0.61^{+0.14}_{-0.15}$  \\
 & 2 & $-1.48^{+0.36}_{-0.36}$& $-1.23^{+0.32}_{-0.31}$ & $0.65^{+0.17}_{-0.16}$ \\ \hline

1RXS J002159.2$-$514028 Sec.\ 2 & 1 & $-1.72^{+0.44}_{-0.42}$& $-1.96^{+0.47}_{-0.47}$ & $0.81^{+0.22}_{-0.20}$  \\
 & 2 & $-1.57^{+0.39}_{-0.39}$& $-0.93^{+0.23}_{-0.24}$ & $0.62^{+0.16}_{-0.16}$ \\ \hline

 QSO B0537$-$441 Sec. 32 \& 33 & 1 & $-1.54^{+0.38}_{-0.37}$& $-1.44^{+0.36}_{-0.36}$ & $0.47^{+0.12}_{-0.11}$  \\
 & 2 & $-1.55^{+0.40}_{-0.41}$& $-1.44^{+0.37}_{-0.36}$ & $0.57^{+0.15}_{-0.14}$ \\ 
 & 3 & $-2.01^{+0.46}_{-0.50}$& $-1.69^{+0.41}_{-0.43}$ & $0.74^{+0.18}_{-0.19}$ \\
 & 4 & $-2.15^{+0.55}_{-0.55}$& $-1.12^{+0.27}_{-0.30}$ & $0.62^{+0.16}_{-0.15}$ \\\hline
\end{tabular}
\end{table*}

In Fig.~3(a), our timing analyses for the 27-day sector 2 observation of 1RXS J002159.2$-$514028 are shown. A peak of around 4.5 days is found which is at least 3$\sigma$ in both the GLSP and time-averaged WWZ results using both simple and bending power laws as underlying red noise models. This 4.5-day signal is found to be stronger in the first half of the observation, as displayed in the WWZ colour diagram so the claim of a QPO for this source is less strong. From the time-averaged WWZ plot one could conclude that the peak at around 11 days is more significant than that at 4.5 days. However, the period of 11 days corresponds to only 2 cycles for the 25-day observation which is not enough to be considered as quasi-periodic. We restrict our analysis in the frequency range of 0.1--2.0 day$s^{-1}$ because the peaks with frequency less than 0.1 day$s^{-1}$ corresponds to a period of more than 10 days which would not be enough for a QPO to be convincing. 

Fig.~3(b) shows the results of the combination of consecutive observations of QSO B0537$-$441 during sectors 32 and 33. 
So, unlike other \textsl{TESS} light curves, this one has a duration  of 53 days. In the GLSP plot, the signal at 6.5 days is found to be at least 3$\sigma$ significant using both red noise models which is also confirmed with time-averaged WWZ analysis. The colour density WWZ plot shows long-lived strong concentrations of power around  6.5 and 11 days. The tentative QPO of around 6.5 days is supported by both methods with at least 3$\sigma$ significance. 
\begin{figure*}
\centering
\includegraphics[scale=0.35]{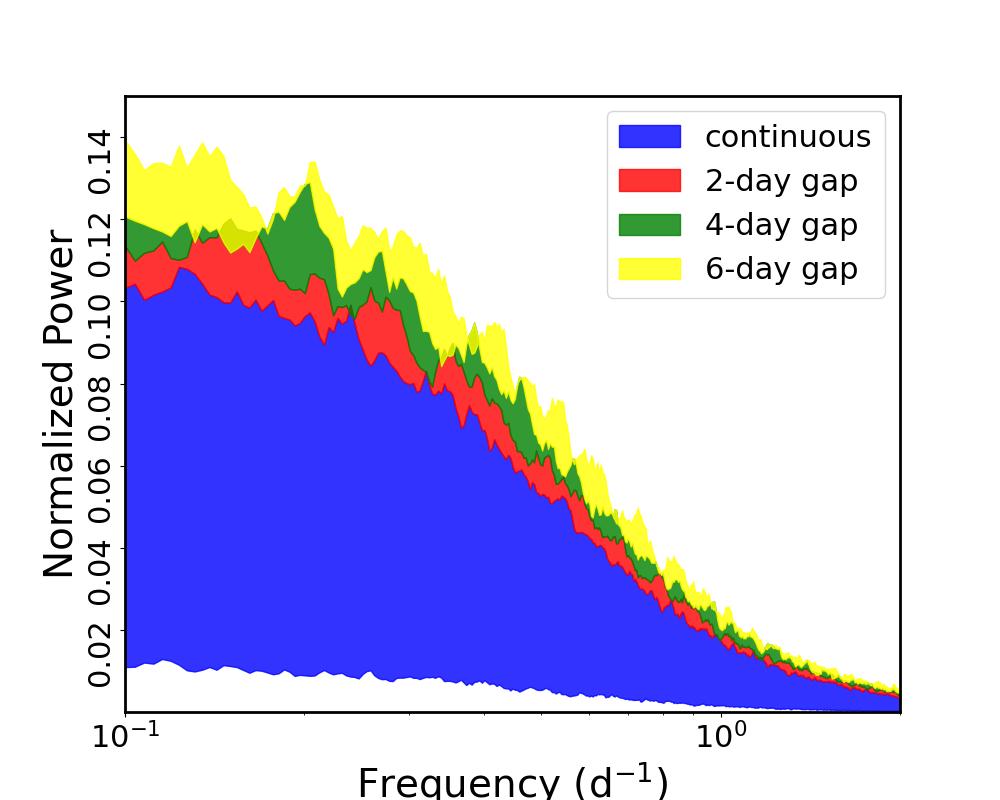}\includegraphics[scale=0.35]{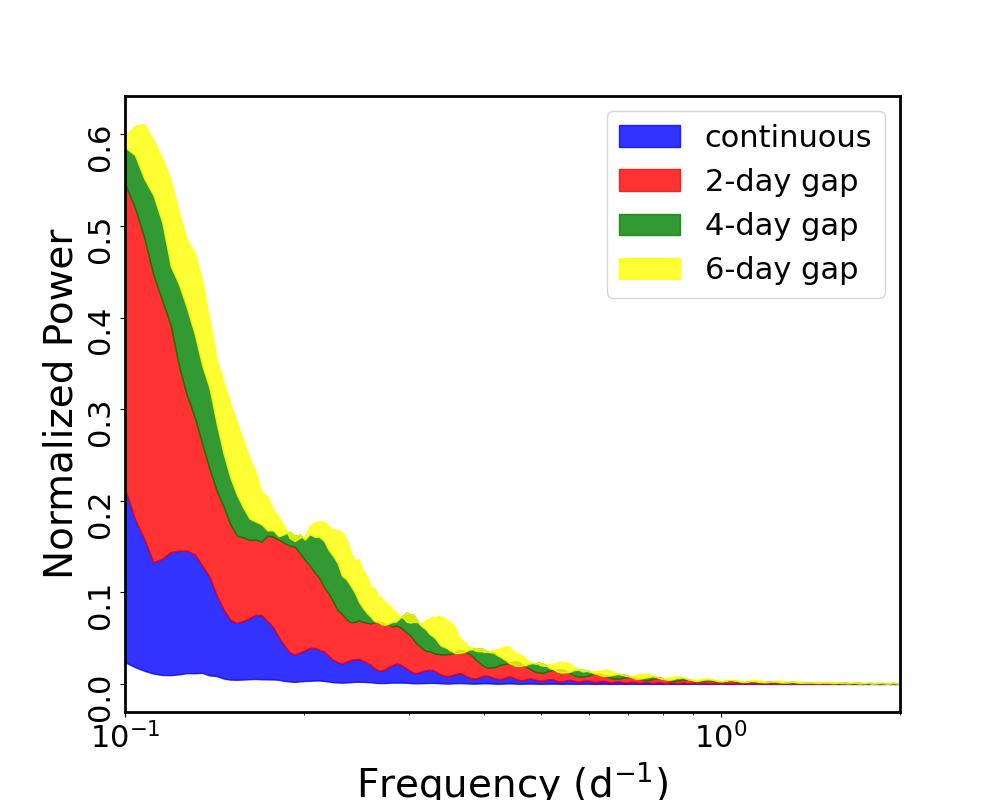}\\

\caption{1$\sigma$ error bounds for the periodogram estimated for different durations of gaps in the middle of observation assuming a bending power law (left) and a simple power law (right) as the underlying red noise model for simulating light curves. }
\end{figure*}\label{fig:sim}

\subsection*{Effects of gaps on periodogram}

The light curves used in this work contain a gap between the sectors due to the orbital motion of the instrument during which the instrument points toward Earth to deliver data. The duration of this gap varies from 1 to 5 days depending on the pointing anomalies associated with the observation and on the cadence masking during the light curve extraction from {\tt quaver} which essentially removes the noise from the data. These gaps are taken into account in the calculation of the significance of the peaks in the GLSP and WWZ analyses. Here we will quantitatively estimate the effect of these gaps in the error estimate of the periodogram.
We first simulate 10,000 26-day light curves with no gaps with the methods explained in the previous section and then estimate the periodogram and the associated 1$\sigma$ confidence interval. In these simulations, we assume both simple power law and bending power law as underlying red-noise models. We repeat these simulations for the light curves with a gap of 2, 4, and 6 days in the middle of the observation. Fig.\ 4 shows the 1-$\sigma$ error associated with the periodogram for different durations of gaps in the light curves that are simulated assuming both a simple power law and a bending power law. For both underlying red noise models, the error estimates increase as the duration of the gap increases. It implies that the significance of the peaks calculated for the light curves in this work may be overestimated but the significance likely would have been greater if the gaps were absent or at least shorter, than those present in these light curves.

\section{Discussion and Conclusions}\label{sec:3}

The majority of the observed emission from blazars is contributed by non-thermal processes in jets and typically substantially exceeds the total emission from the accretion disc and host galaxy \citep[e.g.][]{2001ApJS..134..181J}. This is due to the special relativistic effect of Doppler boosting  produced by the close  alignment of the jet to the line of sight of the observer \citep{1995PASP..107..803U}. These dominant relativistic enhancements in jets manifest themselves in the light curve of blazars as strong variability  occurring on shortened timescales and can be observed throughout the electromagnetic spectrum, from radio to $\gamma$-rays; these variability timescales range from days \citep[e.g.][]{2016ApJ...824L..20A, 2017A&A...603A..25A}  or weeks to even years \citep[e.g.][]{2010MNRAS.402.2087V, 2014APh....54....1A}.  Although the temporal power spectra of these blazar variations are dominated by red noise, a small fraction of them  have QPOs that are statistically significant \citep[see][and references therein]{1998ApJ...504L..71H, 2013MNRAS.436L.114K, 2015Natur.518...74G, 2022Natur.609..265J, 2023T}. 

Several processes can explain  the apparent QPOs with periods of the order of a few days  such as we have tentatively found here. Such periodicity could be related to transient oscillations occurring in various localized regions of a turbulent accretion disc \citep{2004ApJ...609L..63A}  that are advected into the jet. \citet{1997ApJ...476..589P} shows full relativistic calculations for adiabatic oscillations confined in inner regions of the accretion disc due to non-Newtonian gravitational effects of the black hole.  QPOs can also be produced naturally by jet precession that produces periodic changes in Doppler boosting \citep[e.g.][]{1992A&A...259..109G}. Such jet precession could arise from by Lens-Thirring precession of the accretion disc \citep{1998ApJ...492L..59S}. However, these jet precession models are better fits for QPOs with periods of several years. Since the periods of QPOs reported in this work are of the order of a few days, any jet precession model can probably be excluded.

One very attractive model for such quasi-periodic features that is completely internal to the jet involves the development of a kink instability \citep[e.g.][]{2020MNRAS.494.1817D}. The twisting of the toroidal magnetic field and resulting particle acceleration leads to the formation of a kink which is quasi-periodic in distance along the jet. This periodic kink evolves with time in the region of compressed plasma in the magnetized jet and could be detected as a quasi-periodic signature with a period of a few days to weeks \citep[see][and references therein]{2017MNRAS.469.4957B, 2022Natur.609..265J}. 
The transverse motion of the kinked region can be used to estimate the growth time of the kink which can drive such quasi-periodic signatures.  The kink growth time ($\tau_{KI}$) can be calculated as the ratio of transverse displacement of the jet from its center ($R_{KI}$) and average velocity of motion ($\left< v_{tr} \right>$) \citep{2020MNRAS.494.1817D}. The  period in the observer's frame, ($T_{obs}$), of a QPO developed in the kink region,   is
\begin{equation}
T_{obs} = \frac{ R_{KI} }{ \left< v_{tr} \right>\delta },
\end{equation}\label{eq}
\noindent where  $\delta$ is the Doppler factor, $R_{KI}$ can be taken as $10^{16}$--$10^{17}$ cm, as it is the size of the emission region of a typical blazar and $\left< v_{tr} \right>$ can be taken as $\approx 0.16$c as derived in \citet{2020MNRAS.494.1817D}. For $\delta$ =15, $T_{obs}$ is found to be 1--10 days which is consistent with the QPOs claimed for the observations analysed in this work.  A discussion of our MCMC calculations for these blazars is in the Appendix. We find  results that are consistent with the hypothesis that these quasi-periods are due to the growth of kink instabilities instigated in the relativistic jets.

Although \textsl{TESS} solves a large number of problems with ground-based observations, such as diurnal and seasonal monitoring gaps and atmospheric interference with photometric precision, it has a number of properties that make probing AGN variability challenging. The first is its bright limiting magnitude, which rules out a large number of sources in the sky. The second is the short 27-day monitoring baseline for the majority of the sky, and the flux discontinuity across sectors for longer baselines. These preclude robust detection of periods or characteristic timescales at $\gtrsim 5$~days if approximately five cycles are required for confident identification. Improved simultaneous ground-based monitoring in \textsl{TESS}'s future cycles that enable confident inter-sector stitching, or different \textsl{TESS} observation patterns that result in longer sectors, may alleviate this in the future, and expand the parameter space of high-energy phenomena within the grasp of the \textsl{TESS} mission.

\section*{Acknowledgements}

 This work is supported by NASA grant number 80NSSC22K0741. KLS gratefully acknowledges the staff of the K2 Guest Observer office at NASA Ames, especially Christina Hedges, for their assistance and advice in adapting the matrix regression methods.  

 \section*{Data Availability}
The \textsl{TESS} data presented in this paper were obtained from the Mikulski Archive for Space Telescopes (MAST) at the Space Telescope Science Institute.

\appendix
\counterwithin{figure}{section}
\counterwithin{table}{section}

{}

\section{MCMC simulations for the kink instability model}\label{app:2}
\begin{figure*}
\centering
\includegraphics[scale=0.39]{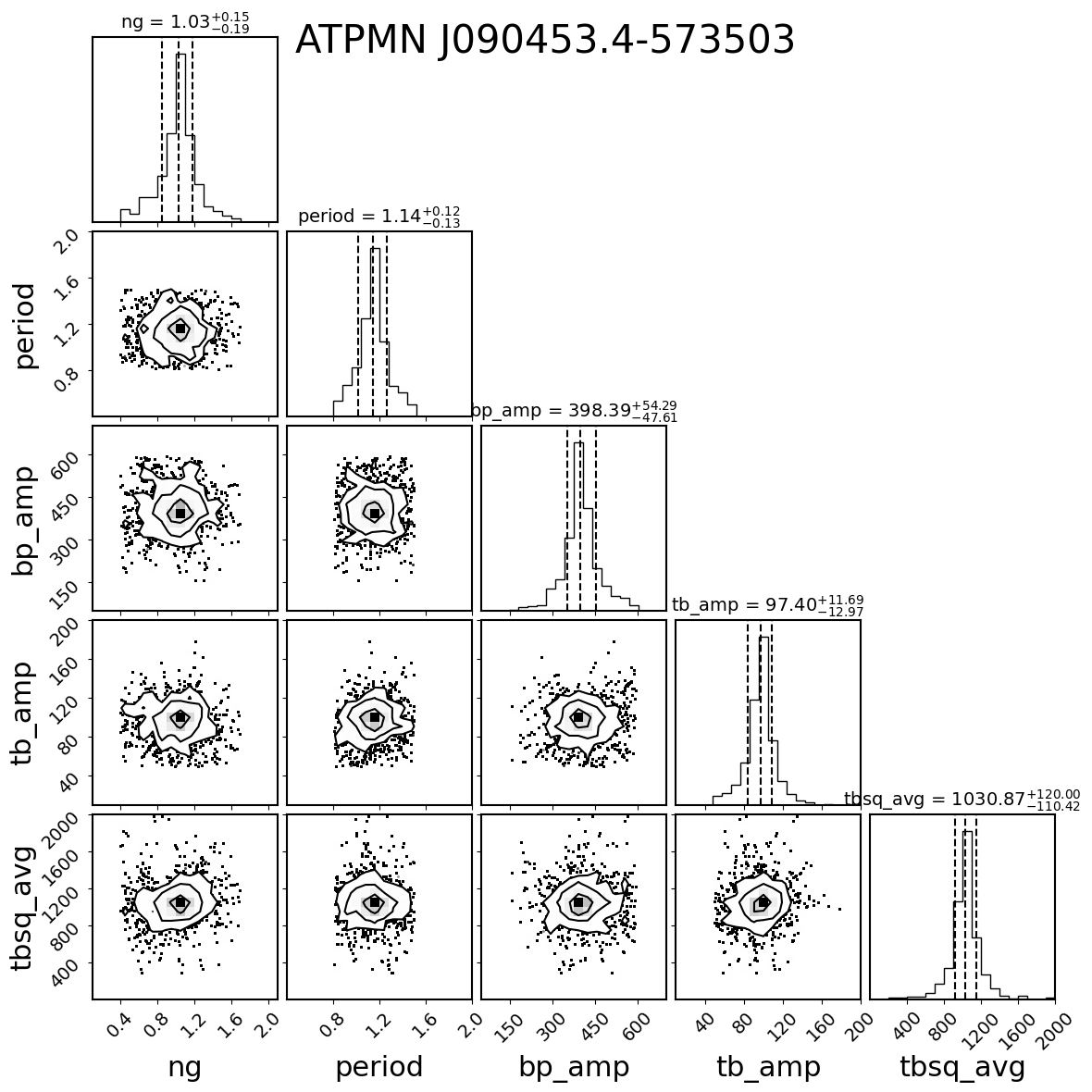}\\
\includegraphics[scale=0.39]{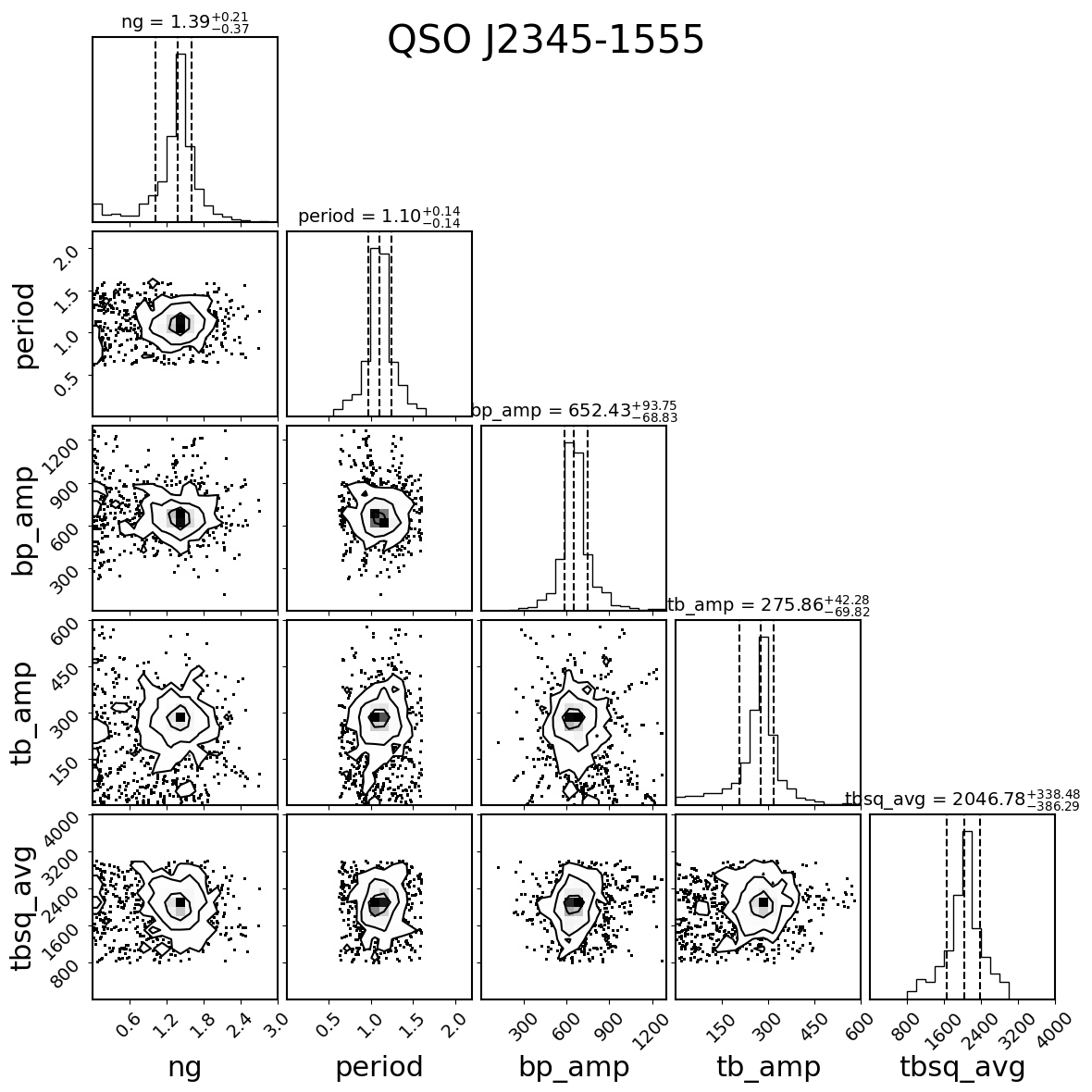}\\
\caption{Corner plots for the posterior distribution of parameters, related to the kink instability model, sampled using 100,000 iterations and 100 walkers through MCMC simulations. Each panel is labelled with the AGN name (and observation Sector numbers for B0422+004).}
\end{figure*}\label{fig6}

\begin{figure*}
\centering
\includegraphics[scale=0.39]{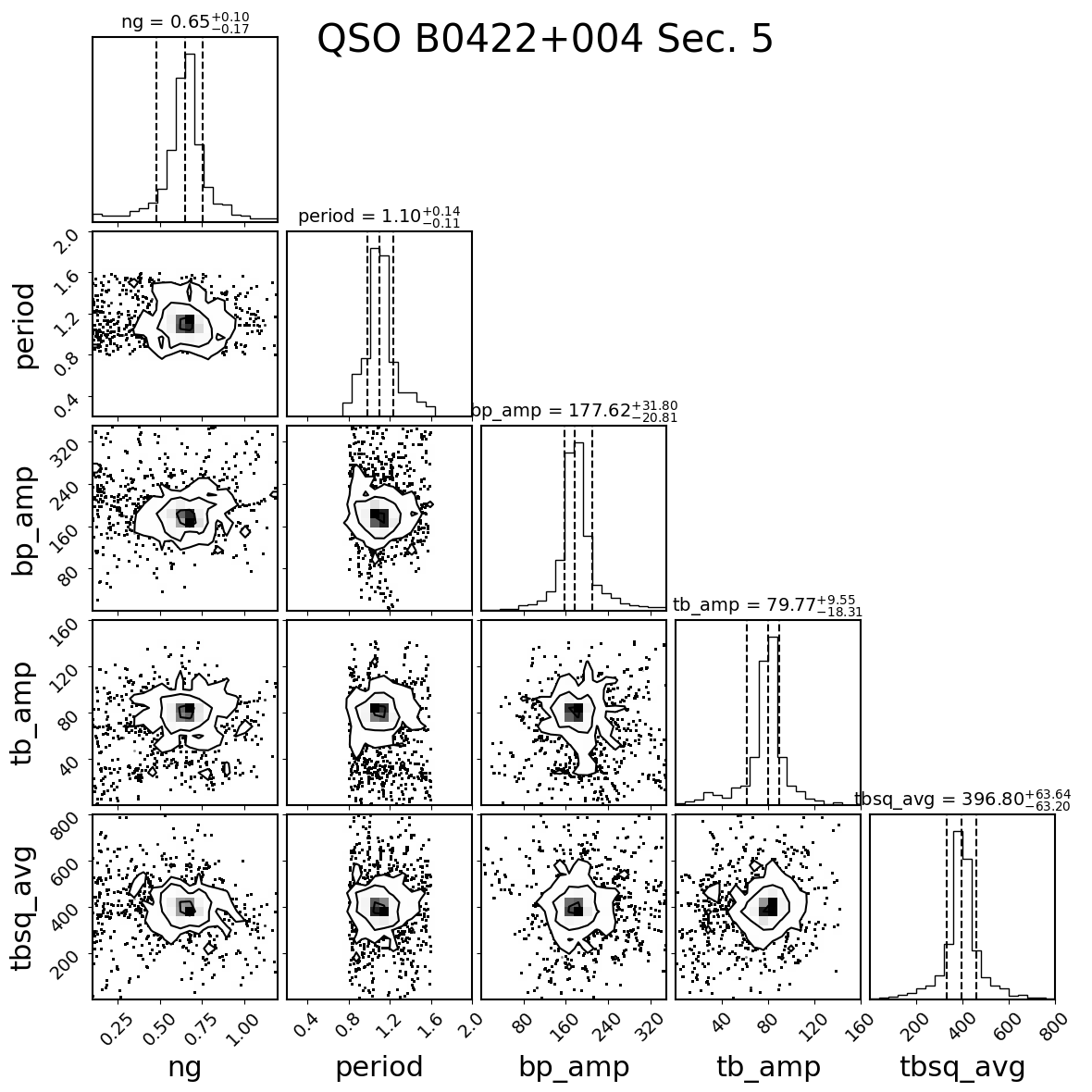}\\
\includegraphics[scale=0.39]{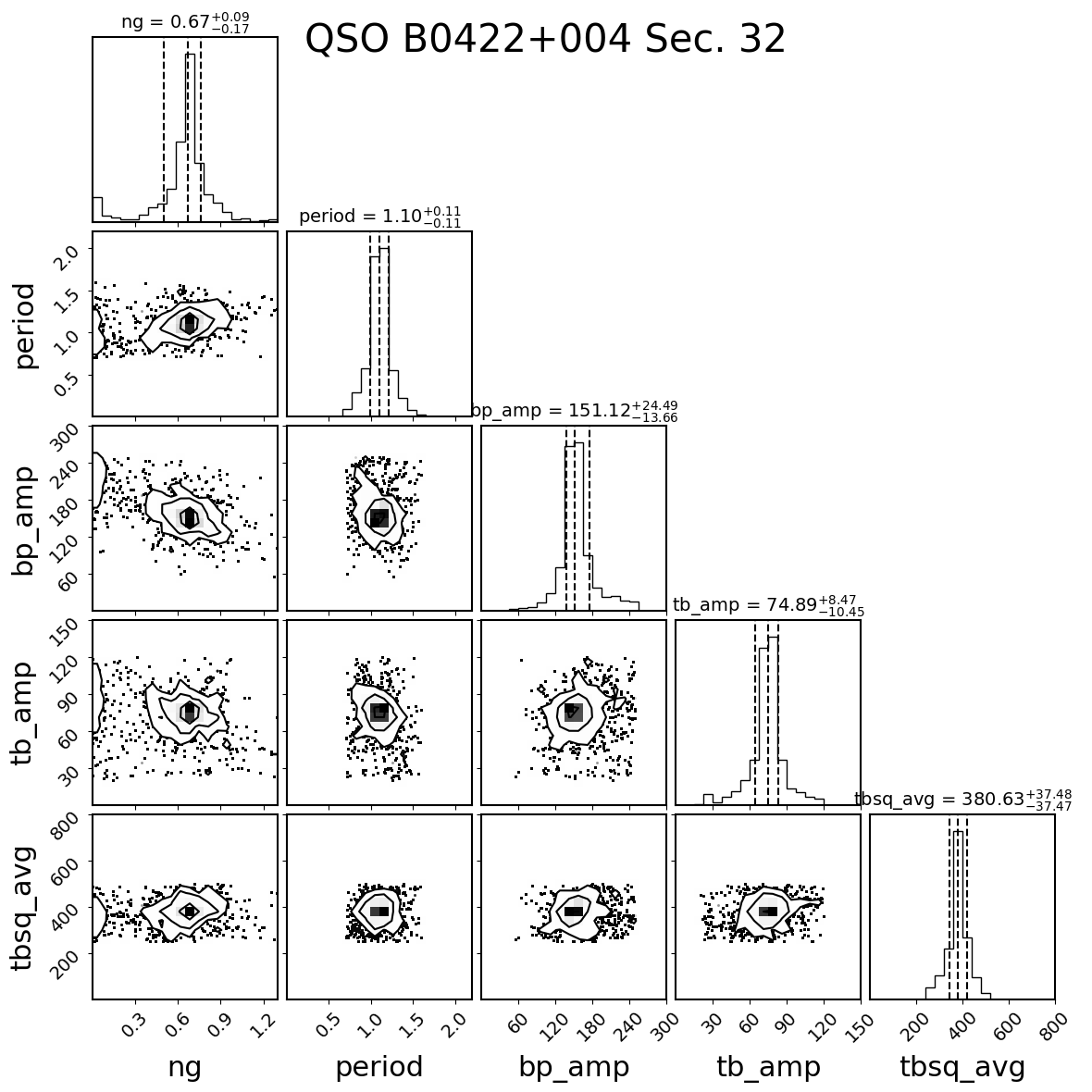}
\caption{As in Fig.\ A1}
\end{figure*}\label{fig7}

\begin{figure*}
\centering
\includegraphics[scale=0.39]{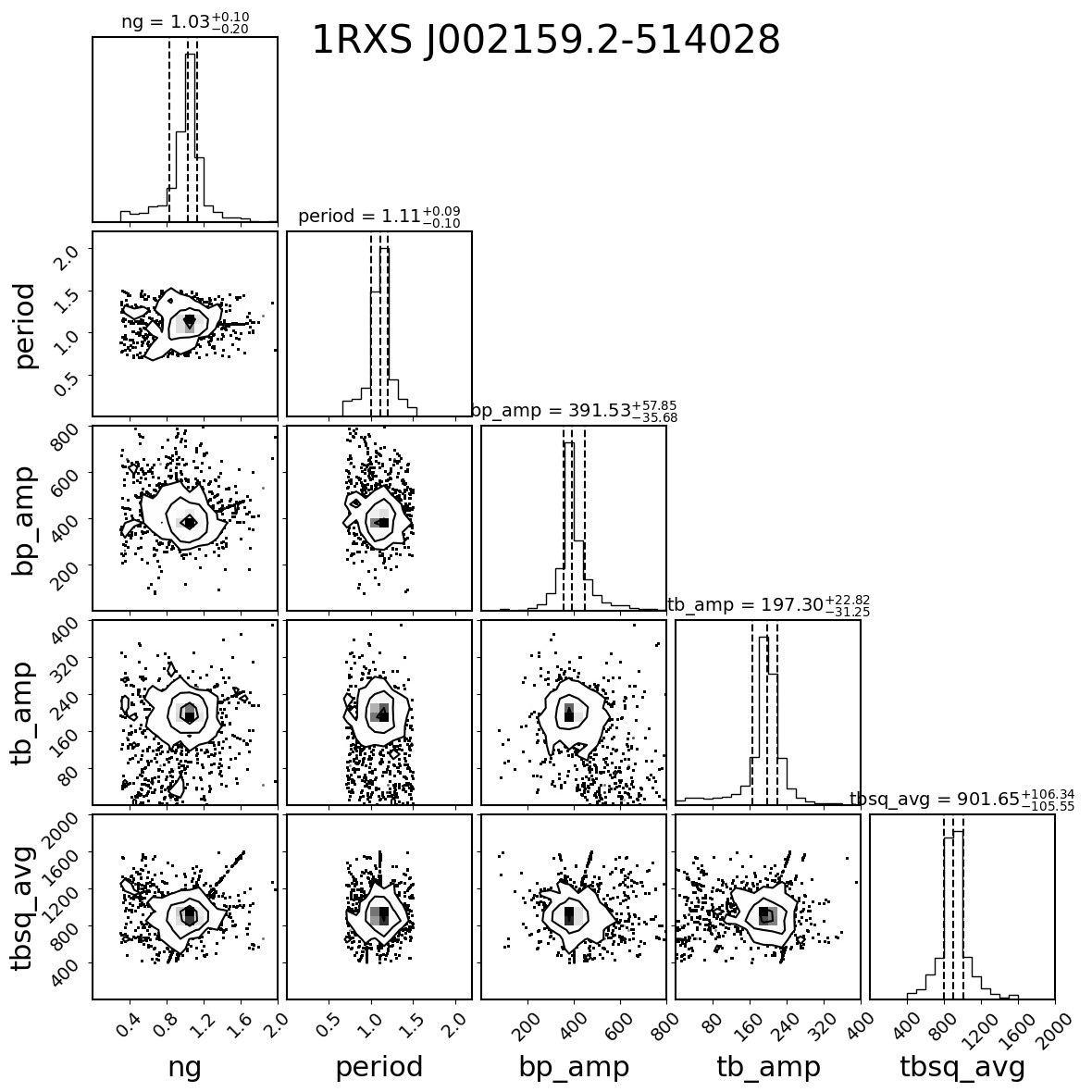}\\
\includegraphics[scale=0.39]{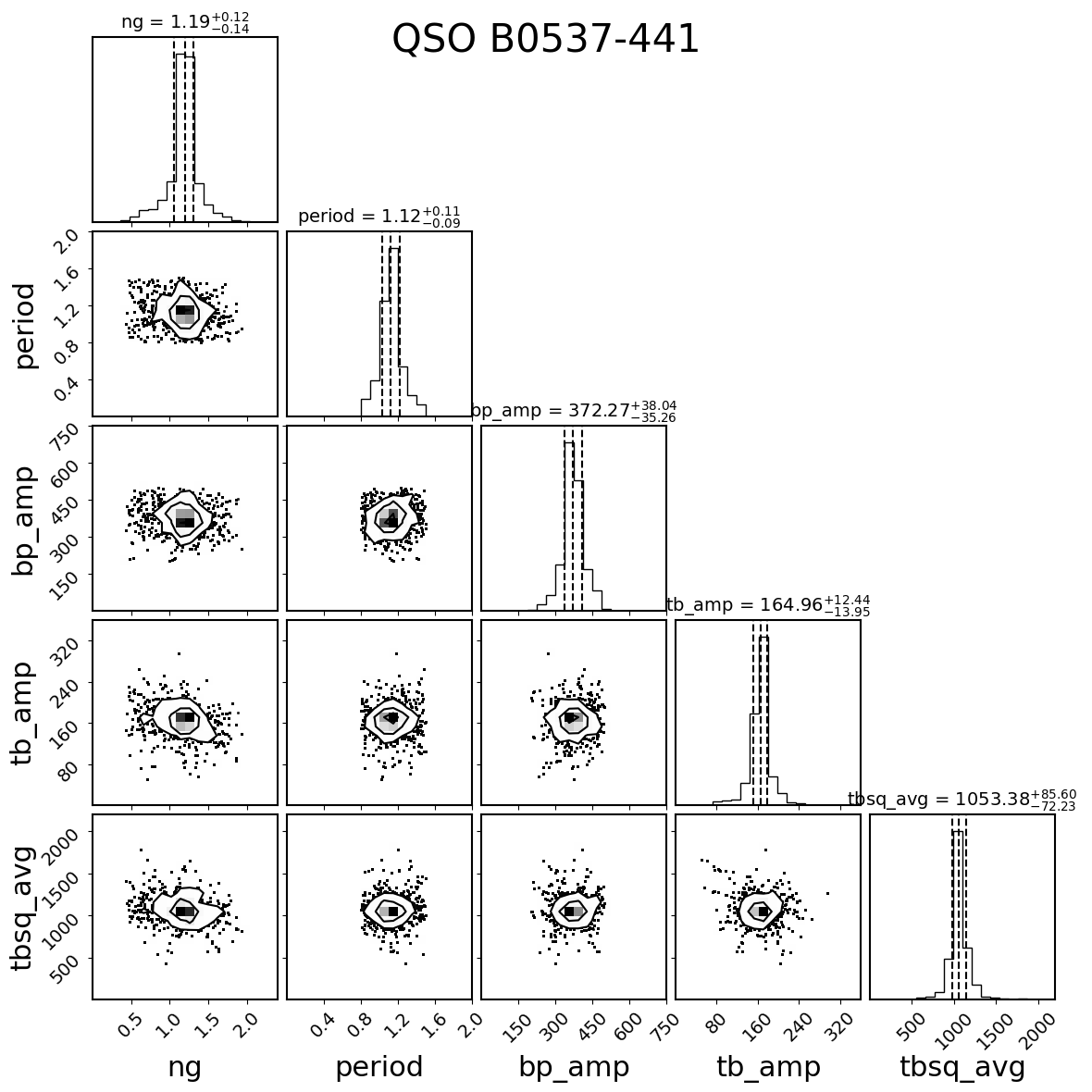}
\caption{As in Fig.\ A1}
\end{figure*}\label{fig8}

We have followed the approach used in \citet{2022Natur.609..265J} and fit the observations with the kink instability model using the publicly available code\footnote{\url{https://zenodo.org/record/6562290}} through MCMC simulations. The code models the fluctuations causing kink instability using the following components; a turbulence profile with amplitude $tb_{amp}$ averaging to $tbsq_{avg}$; average emission power $n_g$; and a sinusoidal toroidal component with amplitude $bp_{amp}$ and period $T$. We used  100,000 iterations with 500 walkers  in the MCMC simulations.  The  posterior distributions of  the 5 model parameters are shown in Figs.\ A1 and A2 as corner plots. The period corresponds to the period of the quasi-periodic kink that originated in the jet. Thus the  QPOs of the order of days found in this work and in \citet{2023T} can reasonably be associated with the temporal growth of this kink. The kink period is found to be in the range of 1.09--1.15 days for all the light curves, which is in agreement with the kink period found in  \citet{2022Natur.609..265J} for BL Lac.


\begin{thebibliography}{}

\bibitem[Abdollahi et al.(2020)]{2020ApJS..247...33A} Abdollahi, S., Acero, F., Ackermann, M., et al.\ 2020, \apjs, 247, 33. doi:10.3847/1538-4365/ab6bcb


\bibitem[Abramowicz et al.(2004)]{2004ApJ...609L..63A} Abramowicz, M.~A., Klu{\'z}niak, W., McClintock, J.~E., et al.\ 2004, \apjl, 609, L63. doi:10.1086/422810

\bibitem[Acciari et al.(2014)]{2014APh....54....1A} Acciari, V.~A., Arlen, T., Aune, T., et al.\ 2014, Astroparticle Physics, 54, 1. doi:10.1016/j.astropartphys.2013.10.004

\bibitem[Ackermann et al.(2016)]{2016ApJ...824L..20A} Ackermann, M., Anantua, R., Asano, K., et al.\ 2016, \apjl, 824, L20. doi:10.3847/2041-8205/824/2/L20

\bibitem[Ahnen et al.(2017)]{2017A&A...603A..25A} Ahnen, M.~L., Ansoldi, S., Antonelli, L.~A., et al.\ 2017, \aap, 603, A25. doi:10.1051/0004-6361/201630347


\bibitem[Astropy Collaboration et al.(2022)]{2022ApJ...935..167A} Astropy Collaboration, Price-Whelan, A.~M., Lim, P.~L., et al.\ 2022, \apj, 935, 167. doi:10.3847/1538-4357/ac7c74

\bibitem[Barniol Duran et al.(2017)]{2017MNRAS.469.4957B} Barniol Duran, R., Tchekhovskoy, A., \& Giannios, D.\ 2017, \mnras, 469, 4957. doi:10.1093/mnras/stx1165



\bibitem[Brasseur et al.(2019)]{2019ASPC..523..397B} Brasseur, C.~E., Phillip, C., Hargis, J., et al.\ 2019, Astronomical Data Analysis Software and Systems XXVII, 523, 397


\bibitem[{\v{C}}emelji{\'c} et al.(2022)]{2022ApJ...933...55C} {\v{C}}emelji{\'c}, M., Yang, H., Yuan, F., et al.\ 2022, \apj, 933, 55. doi:10.3847/1538-4357/ac70cc

\bibitem[Dong et al.(2020)]{2020MNRAS.494.1817D} Dong, L., Zhang, H., \& Giannios, D.\ 2020, \mnras, 494, 1817. doi:10.1093/mnras/staa773

\bibitem[Espaillat et al.(2008)]{2008ApJ...679..182E} Espaillat, C., Bregman, J., Hughes, P., et al.\ 2008, \apj, 679, 182. doi:10.1086/587023

\bibitem[Foster(1996)]{1996AJ....112.1709F} Foster, G.\ 1996, \aj, 112, 1709. doi:10.1086/118137

\bibitem[Gierli{\'n}ski et al.(2008)]{2008Natur.455..369G} Gierli{\'n}ski, M., Middleton, M., Ward, M., et al.\ 2008, \nat, 455, 369. doi:10.1038/nature07277

\bibitem[Gopal-Krishna \& Wiita(1992)]{1992A&A...259..109G} Gopal-Krishna \& Wiita, P.~J.\ 1992, \aap, 259, 109

\bibitem[Graham et al.(2015)]{2015Natur.518...74G} Graham, M.~J., Djorgovski, S.~G., Stern, D., et al.\ 2015, \nat, 518, 74. doi:10.1038/nature14143

\bibitem[Gupta et al.(2019)]{2019MNRAS.484.5785G} Gupta, A.~C., Tripathi, A., Wiita, P.~J., et al.\ 2019, \mnras, 484, 5785. doi:10.1093/mnras/stz395

\bibitem[Halpern et al.(2003)]{2003ApJ...585..665H} Halpern, J.~P., Leighly, K.~M., \& Marshall, H.~L.\ 2003, \apj, 585, 665. doi:10.1086/346106

\bibitem[Hayashida et al.(1998)]{1998ApJ...504L..71H} Hayashida, N., Hirasawa, H., Ishikawa, F., et al.\ 1998, \apjl, 504, L71. doi:10.1086/311574

\bibitem[Huppenkothen et al.(2019)]{2019ApJ...881...39H} Huppenkothen, D., Bachetti, M., Stevens, A.~L., et al.\ 2019, \apj, 881, 39. doi:10.3847/1538-4357/ab258d

\bibitem[Ingram et al.(2009)]{2009MNRAS.397L.101I} Ingram, A., Done, C., \& Fragile, P.~C.\ 2009, \mnras, 397, L101. doi:10.1111/j.1745-3933.2009.00693.x

\bibitem[Jorstad et al.(2001)]{2001ApJS..134..181J} Jorstad, S.~G., Marscher, A.~P., Mattox, J.~R., et al.\ 2001, \apjs, 134, 181. doi:10.1086/320858

\bibitem[Jorstad et al.(2022)]{2022Natur.609..265J} Jorstad, S.~G., Marscher, A.~P., Raiteri, C.~M., et al.\ 2022, \nat, 609, 265. doi:10.1038/s41586-022-05038-9

\bibitem[King et al.(2013)]{2013MNRAS.436L.114K} King, O.~G., Hovatta, T., Max-Moerbeck, W., et al.\ 2013, \mnras, 436, L114. doi:10.1093/mnrasl/slt125

\bibitem[Kato(2005)]{2005PASJ...57..699K} Kato, S.\ 2005, \pasj, 57, 699. doi:10.1093/pasj/57.4.699

\bibitem [Kellermann et al.(1989)]{1989AJ.....98.1195K} Kellermann, K.~I., Sramek, R., Schmidt, M., et al.\ 1989, \aj, 98, 1195. doi:10.1086/115207

\bibitem[Lightkurve Collaboration, 2018]{lightkurve} Lightkurve Collaboration, Cardoso, J.~V.~d.~M., Hedges, C., et al. 2018, ``Lightkurve: Kepler and TESS time series
analysis in Python", Astrophysics Source Code Library,
record ascl:1812.013


\bibitem[Lomb(1976)]{1976Ap&SS..39..447L} Lomb, N.~R.\ 1976, \apss, 39, 447. doi:10.1007/BF00648343

\bibitem[Marscher et al.(2008)]{2008Natur.452..966M} Marscher, A.~P., Jorstad, S.~G., D'Arcangelo, F.~D., et al.\ 2008, \nat, 452, 966. doi:10.1038/nature06895


\bibitem[Perez et al.(1997)]{1997ApJ...476..589P} Perez, C.~A., Silbergleit, A.~S., Wagoner, R.~V., et al.\ 1997, \apj, 476, 589. doi:10.1086/303658


\bibitem[Raiteri et al.(2021)]{2021MNRAS.501.1100R} Raiteri, C.~M., Villata, M., Carosati, D., et al.\ 2021a, \mnras, 501, 1100. doi:10.1093/mnras/staa3561

\bibitem[Remillard \& McClintock(2006)]{2006ARA&A..44...49R} Remillard, R.~A. \& McClintock, J.~E.\ 2006, \araa, 44, 49. doi:10.1146/annurev.astro.44.051905.092532

\bibitem[Ricker (2015)]{2015ESS.....350301R} Ricker, G.~R.\ 2015, AAS/Division for Extreme Solar Systems Abstracts


\bibitem[Scargle(1982)]{1982ApJ...263..835S} Scargle, J.~D.\ 1982, \apj, 263, 835. doi:10.1086/160554


\bibitem[Schulz \& Mudelsee(2002)]{2002CG.....28..421S} Schulz, M. \& Mudelsee, M.\ 2002, Computers and Geosciences, 28, 421. doi:10.1016/S0098-3004(01)00044-9

\bibitem[Smith et al.(2018)]{2018ApJ...860L..10S} Smith, K.~L., Mushotzky, R.~F., Boyd, P.~T., et al.\ 2018, \apjl, 860, L10. doi:10.3847/2041-8213/aac88c


\bibitem[Smith \& Sartori(2023)]{2023ApJ...958..188S} Smith, K.~L. \& Sartori, L.~F.\ 2023, \apj, 958, 188. doi:10.3847/1538-4357/acff5c

\bibitem[Stassun et al.(2019)]{2019AJ....158..138S} Stassun, K.~G., Oelkers, R.~J., Paegert, M., et al.\ 2019, \aj, 158, 138. doi:10.3847/1538-3881/ab3467

\bibitem[Stella \& Vietri(1998)]{1998ApJ...492L..59S} Stella, L. \& Vietri, M.\ 1998, \apjl, 492, L59. doi:10.1086/311075

\bibitem[Stella et al.(1999)]{1999ApJ...524L..63S} Stella, L., Vietri, M., \& Morsink, S.~M.\ 1999, \apjl, 524, L63. doi:10.1086/312291

\bibitem[Tagger \& Pellat(1999)]{1999A&A...349.1003T} Tagger, M. \& Pellat, R.\ 1999, \aap, 349, 1003

\bibitem[Timmer \& Koenig(1995)]{1995A&A...300..707T} Timmer, J. \& Koenig, M.\ 1995, \aap, 300, 707

\bibitem[Tripathi et al.(2021)]{2021MNRAS.501.5997T} Tripathi, A., Gupta, A.~C., Aller, M.~F., et al.\ 2021, \mnras, 501, 5997. doi:10.1093/mnras/stab058

\bibitem[Tripathi et al.(2023)]{2023T} Tripathi, A., Smith, K.~L.,  Wiita, P.~J., \& Wagoner, R.~V.\ 2023, Accepted to MNRAS.

\bibitem[Urry \& Padovani(1995)]{1995PASP..107..803U} Urry, C.~M. \& Padovani, P.\ 1995, \pasp, 107, 803. doi:10.1086/133630

\bibitem[Vaughan(2005)]{2005A&A...431..391V} Vaughan, S.\ 2005, \aap, 431, 391. doi:10.1051/0004-6361:20041453



\bibitem[Villforth et al.(2010)]{2010MNRAS.402.2087V} Villforth, C., Nilsson, K., Heidt, J., et al.\ 2010, \mnras, 402, 2087. doi:10.1111/j.1365-2966.2009.16133.x

\bibitem[Wagoner et al.(2001)]{2001ApJ...559L..25W} Wagoner, R.~V., Silbergleit, A.~S., \& Ortega-Rodr{\'\i}guez, M.\ 2001, \apjl, 559, L25. doi:10.1086/323655

\bibitem[Weaver et al.(2020)]{2020ApJ...900..137W} Weaver, Z.~R., Williamson, K.~E., Jorstad, S.~G., et al.\ 2020, \apj, 900, 137. doi:10.3847/1538-4357/aba693


\bibitem[Witt \& Schumann(2005)]{2005NPGeo..12..345W} Witt, A. \& Schumann, A.~Y.\ 2005, Nonlinear Processes in Geophysics, 12, 345. doi:10.5194/npg-12-345-2005

\bibitem[Zechmeister \& K{\"u}rster(2009)]{2009A&A...496..577Z} Zechmeister, M. \& K{\"u}rster, M.\ 2009, \aap, 496, 577. doi:10.1051/0004-6361:200811296

\bibitem[Zhang et al.(2017)]{2017ApJ...849....9Z} Zhang, P., Zhang, P.-f., Yan, J.-z., et al.\ 2017, \apj, 849, 9. doi:10.3847/1538-4357/aa8d6e

\end{thebibliography}
\end{document}